\documentclass{article}
\usepackage[dvips]{graphics}
\usepackage{amsfonts}
\usepackage{amsmath}
\usepackage{ams}

\newcommand{\pc}{\mathbb{P}}
\newcommand{\reals}{\mathbb{R}}
\newcommand{\mat}[4]{\left(\begin{array}{cc}#1 & #2 \\ #3 & #4
\end{array}\right)}
\newcommand{\un}{\mathsf{l}}
\newcommand{\cfvec}[4]{\begin{array}{c}#1\\#2\\#3\\#4\end{array}}
\newcommand{\chris}[4]{ {#1}_{#2 \,\, #4}^{\,\, #3} }

\begin{document}

\begin{center}
\textbf{\huge Born-Infeld Kinematics}\\[20pt]
Frederic P. Schuller\\[10pt]
Department of Applied Mathematics and Theoretical Physics
\\Centre for Mathematical Sciences
\\University of Cambridge
\\Wilberforce Road
\\Cambridge CB3 0WA
\\United Kingdom
\\E-mail: F.P.Schuller@damtp.cam.ac.uk\\[30pt]

\textbf{Abstract}
\begin{quote}
We encode dynamical symmetries of Born-Infeld theory in a geometry
on the tangent bundle of generally curved spacetime manifolds. The
resulting covariant formulation of a maximal acceleration extension of
special and general relativity is put to use in the discussion of
particular point particle dynamics and the transition to a first
quantized theory.  \\[20pt]
\textit{Keywords:} 
Born-Infeld, pseudo-complex manifold, non-commutative geometry,
maximal acceleration, relativistic phase space, anti-com\-mutation relations.\\[20pt] \textit{Journal Ref.:} Annals of Physics \textbf{299}, 1-34 (2002) 
\end{quote}

\end{center}

%\begin{subject}[Insert header for classifications]
%Use only if your journal has a subject classification requirement
%\end{subject}
\newpage
\tableofcontents
\newpage
\section{Introduction}
Over the last two decades, there has been some interest in and
speculation on the existence of a finite upper bound on accelerations
in fundamental physics, motivated from results in quantum field theory
\cite{qftMA} and string theory \cite{stringMA}. Both of these
theories build on special relativity as a kinematical
framework, but upon quantization, a finite upper bound on
accelerations apparently enters through the back door \cite{Brandt2}.\\
There are interesting results on the promise that the inclusion of a
maximal acceleration on the classical level already, will
positively modify the convergence behaviour of loop diagrams in
quantum field theory \cite{Nesterenko}. These calculations, however, use an ad-hoc
introduction of a maximal acceleration and the authors point out that a
rigorous check would require a consistent classical framework which their
approach is lacking.\\

In this paper, we present such a maximal acceleration extension of
special \textsl{and} general relativity, obtained by 'kinematization' of
\textsl{dynamical} symmetries of the Born-Infeld action, as explained
in the next section.\\

This leads to a non-trivial lift of special and general relativity to
the tangent bundle, or equivalently, the cotangent bundle, of
spacetime.
Born \cite{Born}, among others, remarked that in contrast to our deep
understanding of non-relativistic mechanics in the Hamiltonian
formulation on phase space, special and general relativity are
formulated on spacetime only, and the structure of the associated
phase space is only poorly understood and thus little used. Clearly,
this is likewise true for all dynamical theories based on the
kinematical framework of special and general relativity, most notably
quantum field theory and string theory. In view of the importance of
the non-relativistic phase space structure for the transition to
quantum theory, Born regards a phase space formulation of general
relativity as a necessary step towards a reconciliation of gravity and
quantum mechanics on a descriptional level.\\
Following the observation of Caianiello \cite{Caianiello1981} that the
(co-)tangent bundle of spacetime might be an appropriate stage for a
maximal acceleration extension of special relativity, several
attempts have been made to equip the tangent bundle with a complex
structure, both in the case of flat \cite{Low} and curved
\cite{Brandt} spacetime. These approaches are all inconsistent, as we
will show in section \ref{sec_nogo} that a complex structure is
\textsl{incompatible} with the metric geometry needed to impose an
upper bound on accelerations.\\

Anticipating this negative result, we develop the theory of
pseudo-complex modules and later generalize to pseudo-complex
manifolds. This is then demonstrated to provide the appropriate phase
space geometry circumventing the above mentioned no-go theorem.
We obtain a lift of the Einstein field equations to the tangent
bundle, thus enabling us to formulate a theory of gravity with finite
upper bound on accelerations due to non-gravitational forces. 
Spacetime concepts are regarded as derived ones, and indeed the
existence of a finite maximal acceleration is seen to give rise to a
non-commutative geometry on spacetime, which becomes commutative again
in the limit of infinite maximal acceleration.\\
In dynamical theories featuring a maximal acceleration, second order
derivatives seemed an inconvenient but necessary evil
\cite{Nestsecond, Nestaction} in order to
dynamically enforce the maximal acceleration. 
Exploiting pseudo-complexification techniques, however, we achieve to
recast Lagrangians with second order derivatives into first order form. Rather than
being just a notational trick, the dynamical information on the
maximal acceleration is absorbed into the extended kinematics. This is
shown in detail in section \ref{sec_pointparticle}.\\
 
Putting the full machinery for generally curved pseudo-complex
manifolds to work allows a rigorous discussion of a Kaluza-Klein induced
coupling of a submaximally accelerated particle to Born-Infeld
theory.\\

The last chapter presents a thought on the implications of the
pseudo-complex phase spacetime structure for the transition to quantum theory. One
lesson there is that the pseudo-complex structure \textsl{embraces}
the complex structure, rather than being in opposition to
it.

\section{Dynamics Goes Kinematics}\label{sec_dyngoeskin}
Prior to Einstein's formulation of special relativity, the (at the time)
peculiar role of the speed of light in electrodynamics was regarded a
consequence of the particular dynamics given by Maxwell theory,
\begin{equation}\label{Maxwell}
  \mathcal{L}_{\textrm{Maxwell}} = - \frac{1}{4} F_{\mu\nu} F^{\mu\nu}.
\end{equation}
In modern parlance, one would say that the boost-invariance of
(\ref{Maxwell}) was considered to be a \textsl{dynamical symmetry} of this particular
theory, not necessarily present in other fundamental theories of
nature.\\
Einstein's idea, however, was to regard the symmetries of Maxwell
theory as \textsl{kinematical symmetries}, i.e. due to the structure
of the underlying spacetime, and hence necessarily present in any
theory acting on this stage. Convinced of the correctness of special
relativity, we think today that all sensible dynamics must be
Poincar\'e covariant.\\
Born-Infeld electrodynamics \cite{BornInfeld} 
\begin{equation}\label{BIaction}
  \mathcal{L}_{\textrm{BI}} = \det\left(\eta + b F\right)^\frac{1}{2},
\end{equation}
although indistinguishable from Maxwell theory at
large\footnote{the scale given by the parameter b} distances, modifies
its short range behaviour essentially. 
In particular, Born-Infeld theory features a maximal electric field
strength
\begin{equation}
  E_{\textrm{max}} = b^{-1},
\end{equation}
controlled by the parameter $b$. Hence, coupling a massive particle of
charge $e$ to Born-Infeld theory, 
\begin{equation}\label{prov_BI}
 \mathcal{L} = \mathcal{L}_{\textrm{BI}} + \mathcal{L}_{\textrm{free
  particle}} + \mathcal{L}_{int},
\end{equation}
where
\begin{eqnarray}
 \mathcal{L}_{\textrm{free particle}} &=& p^\mu \dot x^\mu -
  \frac{1}{2} \lambda \left( p_\mu p^\mu - m^2 \right),\\
  \mathcal{L}_{int} &=& - e \dot x^\mu A^\mu,
\end{eqnarray}
we see from the equations of motion that such a particle can at most
experience an acceleration 
\begin{equation}\label{maxBIacc}
  \mathfrak{a} = \frac{e b^{-1}}{m}.
\end{equation}
Clearly, this is a feature solely due to the particular dynamics of
the model (\ref{prov_BI}). But one may well ask whether one can redo
Einstein's trick and convert the \textsl{dynamical} feature of a maximal
acceleration into a \textsl{kinematical} one. Taking the resulting framework
seriously, viable dynamics are then required to be covariant under the
induced symmetry group, which will turn out to include the Lorentz
group.\\
In order to make an \textsl{educated guess} of how the kinematization of the
Born-Infeld symmetries could be achieved, consider the Born-Infeld Lagrangian  
\begin{equation}
\mathcal{L} = {\det}^{1/2}\left(\eta_{\mu\nu} + \frac{e}{m \mathfrak{a}}F_{\mu\nu}\right),
\end{equation} 
where the parameter $b$ is fixed according to relation
(\ref{maxBIacc}). It is fairly easy to verify that this can be
written 
\begin{equation}
  {\det}^{1/4}\left(\eta_{\mu\nu}\right) {\det}^{1/4}\left(\eta_{\mu\nu}
  + \frac{e^2}{m^2 \mathfrak{a}^2} F_{\mu\rho}
  {F_\nu}^\rho\right) = {\det}^{1/4}\left(H_{mn}\right),
\end{equation}
where the we defined the matrix
\begin{equation}
H \equiv \left(\begin{array}{cc} \frac{e}{m \mathfrak{a}}
  F_{\mu\nu} & - \eta_{\mu\nu} \\ \eta_{\mu\nu} & - \frac{e}{m
  \mathfrak{a}} F_{\mu\nu} \end{array}\right),
\end{equation}
and used the determinant identity for $n\times n$ matrices $A, B$,
\begin{equation}
\left|\begin{array}{cc} A & -B \\ B & -A \end{array} \right| =
\left|B\right| \left|B + A B^{-1} A\right|.
\end{equation}
Let $x$ be the worldline of a particle of mass $m$ in spacetime, and $u$ its
four-velocity. Denoting the corresponding curve in 'velocity phase spacetime' 
\begin{equation}
  Y^m \equiv \left(\mathfrak{a} x^\mu, u^\mu\right),
\end{equation}
we recognize that 
\begin{equation}\label{commrel}
  H^{mn} = \frac{i\, m}{\mathfrak{a}}\left[Y^m, Y^n \right]
\end{equation}
corresponds to the canonical commutation relations in presence of an
electromagnetic field
\begin{eqnarray}\label{NCgeo}
  \left[x^\mu, x^\nu\right] &=& - i e m^{-2} \mathfrak{a}^{-2} F^{\mu\nu},\\
  \left[x^\mu, p^\nu\right] &=& - i \eta^{\mu\nu},\\
  \left[p^\mu, p^\nu\right] &=& i e F^{\mu\nu},
\end{eqnarray}
with an additional $\mathfrak{a}^{-2}$-suppressed coordinate
non-commutativity, also controlled by the electromagnetic field
strength tensor. Hence, assuming this \textsl{highly suppressed}
non-commutativity, we have the surprising result
\begin{equation}\label{BIcommform}
  \mathcal{L}_{\textrm{BI}} = {\det}^{1/4} \left(\left[Y^m, Y^n\right]\right), 
\end{equation}
suggesting an encoding of the dynamical symmetries of Born-Infeld
theory in an appropriate geometry of the tangent or cotangent
bundle of Minkowski spacetime.\\

Conventionally, the \textsl{complex}
structure of the cotangent bundle is the key to a geometrical
understanding of Hamiltonian systems \cite{Arnold}, and leads to \textsl{commutation}
relations in the transition to \textsl{bosonic} quantum theory.  
However, in section \ref{sec_nogo}, we show that unless the
underlying spacetime is flat, a complex structure of the associated
phase space is \textsl{incompatible} with a finite upper bound on
accelerations. Hence, even the slightest perturbation of Minkowski
spacetime renders such approaches \cite{Brandt, Low} inconsistent, and hence we
deem them unphysical at all.\\
Fortunately, there is a way round this negative result in form of
equipping phase spacetime with a \textsl{pseudo-complex} structure,
which naturally leads to \textsl{anticommutation} relations for the
associated quantum theory, as shown in section \ref{sec_quant}.
The key mathematical ingredient is the ring of pseudo-complex numbers, and the
module and algebra structures built upon them.
The following chapter is devoted to these mathematical developments. 

\section{Pseudo-Complex Modules \label{sec_pcmodules}}
This section introduces the concept of pseudo-complex numbers,
explores some properties and then focuses on the somewhat subtle
pseudo-complexification of real vectorspaces and Lie algebras.

\subsection{Ring of Pseudo-Complex Numbers}
The pseudo-complex ring is the set
\begin{equation}
  \pc \equiv \left\{a + I b \, |\, a, b \in \reals\right\},
\end{equation}
equipped with addition and multiplication laws induced by those on $\reals$,  
where $I \notin \reals$ is a \textsl{pseudo-complex structure},
i.e. $I^2 = 1$. There is a matrix representation
\begin{equation}
 1 \equiv \mat{1}{0}{0}{1},\qquad\qquad I \equiv \mat{0}{1}{1}{0},
\end{equation}
such that addition and multiplication on $\pc$ are given by matrix
addition and multiplication.
It is easily verified  that $\pc$ is a commutative unit ring with zero divisors
\begin{equation}
  \pc^0_\pm \equiv \left\{ \lambda(1\pm I)\, |\, \lambda\in\reals \right\}.
\end{equation}
$\pc^0_+, \pc^0_- \triangleleft \pc$ are the only multiplicative
ideals in $\pc$, thus they are both maximal ideals. Hence the only
fields one can construct from $\pc$ are 
\begin{eqnarray}
  \pc / \pc^0_+ &\cong& \pc^0_- \cong \reals,\notag\\
  \pc / \pc^0_- &\cong& \pc^0_+ \cong \reals,
\end{eqnarray}
which are too trivial for our purposes. Hence we have to stick to
$\pc$ and deal with its ring structure.\\
For $p=a+Ib \in \pc$, define the pseudo-complex conjugate
\begin{equation}
  \bar p \equiv a - I b.
\end{equation}
The map 
\begin{eqnarray}
  |\cdot|^2 : \pc &\longrightarrow& \reals\notag\\
  |p|^2 &\equiv& p \bar p = a^2-b^2
\end{eqnarray}
is a semi-modulus on the ring $\pc$, which now decomposes into three classes
\begin{equation}
  \pc = \pc^+ \cup \pc^- \cup \pc^0,
\end{equation}
according to the sign of the modulus:
\begin{eqnarray}
  \pc^+ &\equiv& \left\{p \in \pc : |p|^2 > 0 \right\},\notag\\ 
  \pc^- &\equiv& \left\{p \in \pc : |p|^2 < 0 \right\},\\  
  \pc^0 &\equiv& \left\{p \in \pc : |p|^2 = 0 \right\} = \pc^0_\pm.\notag
\end{eqnarray} 
Define the exponential map
\begin{eqnarray}
  \exp: \pc &\longrightarrow& \pc^+\notag\\
     \exp(p) &\equiv& \sum_{n=0}^{\infty} \frac{p^n}{n!}.
\end{eqnarray}
In terms of $p=a+Ib$ this yields
\begin{equation}
\exp(a+Ib) = \exp(a) \left\{\cosh(b) + I \sinh(b) \right\}
\end{equation}
and hence $\exp$ converges on all of $\pc$ and is one-to-one. 
As $\pc$ is commutative, the functional identity
\begin{equation}
  \exp(p) \exp(q) = \exp(p + q) 
\end{equation} holds for all pseudo-complex numbers $p,q \in \pc$.
Note, in particular, that for any $\exp(p)$, there is always a unique multiplicative inverse,
namely $\exp(-p)$.\\
Using the exponential map, we get the 'polar' representations
\begin{eqnarray}
  \pc^+ &=& \left\{ r \exp(I \psi) | r,\psi \in \reals \right\},\notag\\
  \pc^- &=& \left\{ Ir \exp(I \psi) | r,\psi \in \reals \right\},\\
  \pc^0_\pm &=& \left\{ \lambda(1 \pm I) | \lambda \in \reals \right\}.\notag
\end{eqnarray} 
% \begin{figure}[h]
%\epsfig{file=pcn_polar.eps, height=5cm, width=8cm}
% \caption{\label{fig_pcnpolar} polar representations of (a) pseudo-complex
% and (b) complex numbers}
% \end{figure} 
It is easily verified that the symmetry transformations on $\pc$
preserving the semi modulus are the $(1+1)$-dimensional Lorentz group:
\begin{equation}
  O_\pc(1) \cong O_\reals(1,1)
\end{equation}

\subsection{$\pc$-Modules and Lie Algebras}
Usually, the Lie algebras occuring in physical applications are real or
complex vector spaces. However, the most general algebraic definition \cite{Lang} of a Lie
algebra only demands it be a module over a commutative ring. Hence we
can sensibly 
define the pseudo-complex extension $L_\pc$ of a real Lie algebra $L$
by
\begin{equation}
  L_\pc \equiv \left\{t + I s | t, s \in L \right\},
\end{equation}
which is a free $\pc$-module, as $L$ is a vector space. The Lie bracket on
$L_\pc$ is induced by that on $L$. Its multilinearity follows directly
from the commutativity of $\pc$. Clearly, 
$$\dim_\pc(L_\pc) = \dim_\reals(L) = \frac{1}{2} \dim_\reals(L_\pc).$$
Let $T_i$ with $i=1,\dots, \dim_\reals(L)$ be the generators of the
real Lie algebra $L$. Then we have
\begin{equation}
  L_\pc = \left<T_i\right>_\pc = \left<T_i, I T_i\right>_\reals,
\end{equation}
and we will switch between these two pictures where appropriate. As
$\pc$ is commutative, we have
\begin{equation}
  \left[\exp(p_i T_i)\right]^{-1} = \exp(- p_i T_i).
\end{equation}
Hence we can obtain the connection component $G^{id}$ of the associated
Lie group $G$ by exponentiation of the algebra
\begin{equation}
  G^{id} = \exp\left(\pc L\right) = \exp\left(\reals L_\pc\right).
\end{equation}
Let the real vector space $V$ be a representation of the real Lie
group $L$. Then $L_\pc$ acts naturally on the pseudo-complexification
of $V$,
\begin{equation}
  V_\pc \equiv \left\{x + I u | x,u \in V \right\}
\end{equation} 
which is a free $\pc$-module of dimension $\dim_\reals V$. $V_\pc$ also
being an $\reals$-vectorspace of dimension $2 \dim_\reals$, we can identify $V_\pc$ with the tangent bundle
$TV$ via 
\begin{equation}
   V_\pc = \left\{x + I u | x\in V, u \in T_xV \right\}. 
\end{equation}
The bundle projection is then given in this language by  
\begin{eqnarray}\label{bundleproj}
 \pi: TV\equiv V_\pc &\longrightarrow& V \notag\\
  \pi(X) &\equiv& \frac{1}{2}(X + \bar X)
\end{eqnarray}

\subsection{The Pseudo-Complex Lorentz Group\label{sec_PLorentz}}
Let $\eta \equiv \textrm{diag}\left(1,-1,\dots,-1\right)$ have
signature $(1,n-1)$  and consider the
pseudo-complex extension $so_\pc(1,n-1)$ of the real Lorentz algebra
$so_\reals{(1,n-1)}$. Exponentiation gives the pseudo-complex Lorentz group    
\begin{equation}
  SO_{\pc}(1,n-1) \equiv \left\{\Lambda \in
  \textrm{Mat}\left(n, \pc\right) | \Lambda^T \eta \Lambda
  = \eta, det \Lambda = 1 \right\}
\end{equation}
Clearly, for any $U \in T\pc^n$, the expression
\begin{equation}\label{pcmetric}
  \eta\left(U,U\right) \equiv U^\mu U^\nu \eta_{\mu\nu} 
\end{equation}
is invariant under the action of $SO_{\pc}(1,n-1)$. Expanding
(\ref{pcmetric}) using $U^\mu = u^\mu + I a^\mu$, $u,a \in \reals^{r+s}$ yields
\begin{equation}\label{Pnorm}
\eta\left(U,U\right) = \left(u^\mu u_\mu + a^\mu a_\mu\right) + I \left(2 u^\mu
a_\mu\right).
\end{equation}
Let $M^{\mu\nu}$ be the standard generators of the real Lorentz-group,
then 
\begin{equation}
so_\pc(1,n-1) = \left<M^{\mu\nu}, I M^{\mu\nu}\right>_\reals
\end{equation}
and the pseudo-complex linear combinations
\begin{equation}
  G^{\mu\nu} \equiv \frac{1}{2} (M^{\mu\nu} + I M^{\mu\nu}), \quad \bar G^{\mu\nu} \equiv \frac{1}{2}(M^{\mu\nu} - I M^{\mu\nu})
\end{equation}
generate two decoupled real Lorentz algebras, and hence
\begin{equation}
  so_\pc(1,n-1) \cong so_\reals(1,n-1) \oplus so_\reals(1,n-1).  
\end{equation}
Note that the real and pseudoimaginary part of (\ref{Pnorm}) are
preserved separately under the action of $O_\pc(1,n-1)$. Hence,
we can switch between the picture of a \textsl{metric module} and a \textsl{bimetric
vector space}: 
\begin{equation}\label{isomorphy}
  \left(\pc^n, \eta\right) \cong \left(T\reals^n, \eta^D, \eta^H\right),
\end{equation}
where $\eta^D=\eta\otimes 1$ and $\eta^H= \eta\otimes I$ denote the
diagonal and horizontal lifts to the tangent bundle
\cite{Yano1973}, respectively, of the Minkowski metric.

\subsection{The Pseudo-Complex Sequence \label{sec_pciterat}}
Define inductively, for all $n \in \mathbb{N}$,
\begin{eqnarray}
  \pc^{(0)} &\equiv& \reals,\\
  \pc^{(n+1)} &\equiv& \left\{1\otimes a + I\otimes b \, |\,
  a,b\in\pc^{(n)} \right\}, 
\end{eqnarray}
the sequence of pseudo-complex rings. $\pc^{(n)}$ is called the
pseudo-complex ring of rank $n$. Commutativity is easily shown by
induction. Clearly, $\pc^{(n)}$ is a vector
space of dimension $2^n$, with canonical basis 
\begin{equation}
 1 \otimes 1 \otimes \dots \otimes 1, 1 \otimes 1 \otimes
 \dots \otimes I, \dots, I \otimes I \otimes \dots \otimes I.\notag
\end{equation}
Identifying the $b$-th canonical basis vector $\mathbf{b}$ in $\pc^{(n)}$ with a binary
number $0 \leq b \leq 2^n-1$ via $1=0$, $I=\un$, multiplication
between basis elements corresponds to the 'exclusive or' operation $\sqcup$ 
\begin{equation}
\begin{array}{c|cc}
   \sqcup & 0 & \un \\
\hline
   0 &  0 & \un\\
   \un &  \un & 0
\end{array}
 \end{equation}   
Hence the multiplication of two $n$-th rank pseudo-complex numbers
\begin{equation}
  p = \sum_{b=0}^{2^n-1} p_b \,\mathbf{b}\notag, \qquad
  q = \sum_{b=0}^{2^n-1} q_c \,\mathbf{c}\notag
\end{equation}
is given by
\begin{eqnarray}
  p q &=& \sum_{b,c}^{2^n-1} p_b q_c \,\mathbf{b} \sqcup
  \mathbf{c}\notag\\
      &=& \sum_{b,d}^{2^n-1} p_b q_{b \sqcup d} \, \mathbf{d},
\end{eqnarray}
observing that $b \sqcup c = d$ if and only if $b \sqcup d = c$.
We use the extension of this result to the infinite-dimensional case
to \textsl{define} the multiplication law on 
\begin{equation}
  \pc^{(\infty)} \equiv \left\{p: \mathbb{N} \longrightarrow \reals \right\},
\end{equation}
the set of all real sequences.\\
From the binary representation, the sets $\pc^{(n)}$ and
$\pc^{(\infty)}$ are seen to be commutative rings with unit $\mathbf{0}$. 
Hence on can define the $n$-th rank pseudo-complexification of a real Lie algebra $L$
\begin{equation}
  L_{\pc^{(n)}} \equiv \left\{\sum_{b=0}^{2^n-1} t_b \mathbf{b} \,|\,
  t_b \in L \right\}
\end{equation}
which is a free module of dimension
\begin{equation}
  \dim_\pc^{(n)}(L_{\pc^{(n)}}) = \dim_\reals(L) = 2^{-n} \dim_\reals(L_{\pc^{(n)}}), 
\end{equation}
and the $n$-th rank pseudo-complexification of a real vectorspace $V$,
\begin{equation}
  V_{\pc^{(n)}} \equiv  \left\{\sum_{b=0}^{2^n-1} x_b \mathbf{b} \,|\,
  x_b \in V \right\}
\end{equation}
being a free $\pc^{(n)}$-module, or a real vectorspace of dimension
$2^n \dim_\reals(V)$. We can identify $V_{\pc^{(n)}}$ with the $n$-th
tangent bundle $T^nV$, i.e. the bundle of $n$-jets \cite{Yano1973}. 

\section{Maximal Acceleration Extension of Special Relativity}\label{sec_MSR}
We obtain a maximal acceleration extension of special relativity by
pseudo-complexification of Minkowski spacetime
\begin{equation}
  \reals^{1,3} \longrightarrow \pc^{1,3}
\end{equation}
and appropriate
lifts of all spacetime concepts to the resulting module. Alternatively to
pseudo-complexification, all this can be understood in terms of lifts
to the tangent bundle of spacetime, at the cost of having to deal with
two metrics.\\ 
That the geometry obtained in this way indeed encodes the Born-Infeld
kinematics, is shown in section \ref{sec_BIrevisited}.
The theory is formulated entirely independent of spacetime
concepts. This involves the slight paradigm shift of thinking of the
objects defined on pseudo-complexified spacetime as primary, rather
than being induced from the familiar spacetime concepts, which will be
regarded as derived ones. 

\subsection{Mathematical Structure}
We introduce a fundamental constant $\mathfrak{a} $
of dimension \textsl{length${}^{-1}$}, called \textsl{maximal
acceleration}. The stage for extended relativistic physics is
pseudo-complexified Minkowski spacetime 
$$\pc^{1,3} \equiv \left(\pc^4, \eta\right),$$
equipped  with a pseudo-complex valued two-form 
\begin{equation}
  \eta: \pc^{4} \times \pc^{4} \longrightarrow \pc
\end{equation}
of signature $(1,3)$. Note that without further restriction, $\eta$ is
not a real-valued semi-norm on $\pc^{1,3}$, as its pseudo-imaginary
part is non-vanishing in general.\\
The symmetry algebra of $\pc^{1,3}$ is, by construction,
the pseudo-complexified Lorentz algebra 
$$o_\pc(1,3).$$
We introduce a preferred class of coordinate systems, generalizing
the inertial frames of special relativity.
A basis $\left\{e_{(\mu)}\right\}$ of $T\pc^{1,3}$ with 
\begin{equation}
\eta\left(e_{(\mu)}, e_{(\nu)}\right) = \eta_{\mu\nu} \equiv \textrm{diag}\left(1,-1,-1,-1\right)
\end{equation}
is called a \textsl{uniform basis} or \textsl{uniform frame}. Coordinates of $\pc^{1,3}$ given with respect
to such a basis are called \textsl{uniform coordinates}.\\
We will always work in uniform coordinates in this chapter.
Clearly, under the action of $SO_\pc(1,3)$, a uniform basis is
transformed to a uniform basis.\\

Let $x: \reals \longrightarrow \reals^{1,3}$ be a timelike curve in
real Minkowski spacetime. Then the natural lift $X\equiv x^*$ to
pseudo-complexified spacetime 
\begin{eqnarray}
x: \reals &\longrightarrow& \reals^{1,3} \notag\\
  &\downarrow *&  \notag\\
X: \reals &\longrightarrow& \pc^{1,3}
\end{eqnarray} 
is defined by
\begin{equation}\label{natlift}
  X \equiv x^* \equiv \mathfrak{a} x + I u
\end{equation}
where $u\equiv\frac{dx}{d\tau}$ and $d\tau^2\equiv\eta(dx,dx)$.
Let $X: \reals \longrightarrow \pc^{1,3}$ be a
curve in configuration space. 
$X$ is called an \textsl{orbit} iff there exists a uniform frame
$\Sigma$ such that
\begin{equation}
  X = \pi(X)^*,
\end{equation}
where $*$ denotes the natural lift (\ref{natlift}) and $\pi$ the
projection (\ref{bundleproj}). The frame $\Sigma$ is called an
\textsl{inertial frame}.\\
The line element of the projection $\pi(X)$ of an arbitrary orbit $X$
in a particular uniform (not necessarily inertial) frame is denoted by $d\tau$ and given by
\begin{equation}
  d\tau^2 \equiv \eta\left(d\pi(X),d\pi(X)\right).
\end{equation}
Note that this quantity is frame-dependent.\\
It is clear that for an orbit $X$ given in inertial coordinates
$X^\mu = x^\mu + I u^\mu$, we always have the relation $u =
\frac{dx}{d\tau}$. Further it follows that the projection $\pi(X)$ is
necessarily timelike in inertial coordinates. 
Now consider an orbit $X$ in an inertial frame. The orthogonality of $dx$ and
$du=d\frac{dx}{d\tau}$ yields
\begin{equation}\label{eqn_dxdxnull}
  \eta\left(dX, dX\right) \in \reals
\end{equation}
in inertial coordinates, but due to the $SO_\pc(1,3)$-invariance of
(\ref{eqn_dxdxnull}) this result even holds in any uniform frame. 
Hence, along an orbit, the generically $\pc$-valued two-form $\eta$
provides a \textsl{real}-valued semi inner product, allowing the following classification:
An orbit $X$ is called \textsl{submaximally accelerated}, if
\begin{equation}
  d\omega^2 \equiv \left(dX,dX\right) > 0
\end{equation}
everywhere along the orbit. Note that the line element $d\omega$ is an $O_\pc(1,3)$ scalar.
Observe that an orbit $X$ is submaximally accelerated if and only if
the projection $\pi(X)$ has Minkowski curvature $g<\mathfrak{a}$. 
This is seen as follows. In an inertial frame, let $x \equiv \pi(X)$, $u \equiv
\frac{dx}{d\tau}$ and $a \equiv \frac{du}{d\tau}$. Then
$X=\mathfrak{a} x + I u$ and we have from (\ref{Pnorm})
\begin{equation}
  d\omega^2 = \left(1 - \mathfrak{a}^{-2} g^2\right) \mathfrak{a}^2
  d\tau^2
\end{equation}
where $g$ is the Minkowski-scalar acceleration of the trajectory $x$. Hence, 
$$ d\omega^2 > 0 \Leftrightarrow g<\mathfrak{a}$$
as $d\tau^2>0$ as $x=\pi(X)$ is timelike everywhere. As $d\omega^2$ is
$SO_\pc(1,3)$-invariant, the result holds in any uniform frame.\\
This gives us the interpretation of the constant $\mathfrak{a}$ as the
upper bound for accelerations in this theory, and thus justifies the
terminology.\\

In analogy to the notion of \textsl{rapidity} in special relativity,
it is useful to introduce a convenient non-compact measure for
accelerations. This will clear up notation later on. 
Let $X$ be a submaximally accelerated orbit, and $g$ be the
Minkowski curvature of the projection $\pi(X)$ for a particular
uniform observer. Then the \textsl{accelerity} $\alpha$ of the
trajectory is given by
\begin{equation}\label{accelerity}
  \tanh(\alpha) = \frac{g}{\mathfrak{a}}.
\end{equation}  
Hence the relation between the $O_\pc(1,3)$-invariant line element
$d\omega$ of an orbit and the Minkowski line element $d\tau$ of the
projection $\pi(X)$ is 
\begin{equation}
  d\omega = \frac{\mathfrak{a}}{\cosh(\alpha)} d\tau.
\end{equation}
Note that although $d\tau$ and $\alpha$ are frame dependent, the
combination on the right hand side above is manifestly frame
independent, as $d\omega$ is.\\
Finally, we define the \textsl{eight-velocity} $U$ of a submaximally
accelerated orbit $X$ as
\begin{eqnarray}
  U: \reals &\longrightarrow& T\pc^{1,3}\\
  U &\equiv& \frac{dX}{d\omega}.
\end{eqnarray}
This is well-defined due to the $O_\pc(1,3)$-invariance of $d\omega$.
Sometimes we will consider the real and
pseudo-imaginary part of $U$ in uniform coordinates, which
we will denote
\begin{equation}\label{tilde}
  U \equiv \tilde{u} +  I \tilde{a} = \cosh(\alpha) \left(u + I
  \mathfrak{a}^{-1} a \right),
\end{equation}
where $u$ and $a$ are the four-velocity and acceleration of the
corresponding spacetime trajectory $\pi(X)$.

\subsection{Linear Uniform Acceleration in Extended Special
  Relativity \label{sec_uniformMSR}}
\begin{figure}[h]
\includegraphics[0,0][100,140]{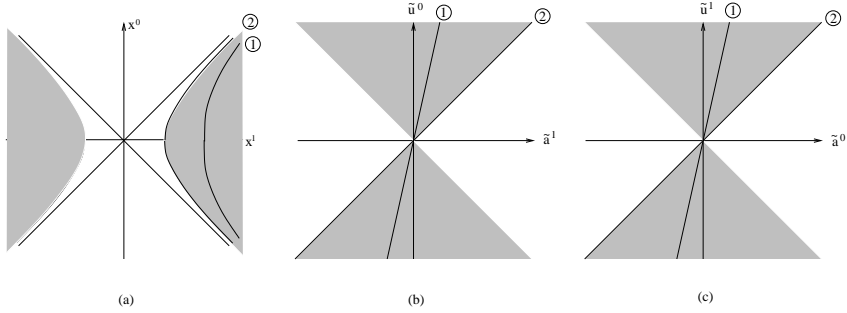}
\caption{\label{fig_msrspectrum} Spectrum of uniformly accelerated
curves in ESR}
\end{figure}

It is instructive to study orbits which project to trajectories of linear
uniform acceleration, i.e. constant accelerity $\alpha$. Let $U$ be
the eight-velocity of such an orbit. From $\eta(U,U)=1$ and using 
(\ref{accelerity}), (\ref{tilde}), we get
\begin{eqnarray}
  \tilde u^\mu \tilde u_\mu &=& 1 + \sinh^2(\alpha),\\
  \tilde u^\mu \tilde a_\mu &=& 0.\label{uaortho}
\end{eqnarray}
Choosing a Lorentz frame such that 
$$u^2=u^3=a^2=a^3=0,$$
(\ref{uaortho}) becomes
\begin{equation}
  \tilde u^0 \tilde a^0 - \tilde u^1 \tilde a^1 = 0,
\end{equation}
which is solved by 
\begin{eqnarray}
  \tilde a^0 &=& \gamma \tilde u^1,\\
  \tilde a^1 &=& \gamma \tilde u^0,
\end{eqnarray}
for some function $\gamma$. Constant accelerity gives
$$\tilde a^\mu \tilde a_\mu = - \cosh^2(\alpha)
  \frac{g^2}{\mathfrak{a}^2} = - \sinh^2(\alpha).$$ 
On the other hand,
$$\tilde a^\mu \tilde a_\mu = - \gamma^2 \tilde u^\mu \tilde u_\mu = -
\gamma^2 \left(1 + \sinh^2(\alpha)\right).$$ 
Hence for linear uniform acceleration of modulus $g$,  
\begin{eqnarray}
  \tilde a^0 &=& \frac{g}{\mathfrak{a}} \tilde u^1, \label{a0is}\\
  \tilde a^1 &=& \frac{g}{\mathfrak{a}} \tilde u^0, \qquad \textrm{
  where } g < \mathfrak{a}.\label{a1is}
\end{eqnarray}
Now consider the projections  
\begin{eqnarray}
  \pi_{01}(\tilde u^0, \tilde u^1, \tilde a^0, \tilde a^1) &=& (\tilde
  u^0, \tilde a^1) \label{pi01},\\
  \pi_{10}(\tilde u^0, \tilde u^1, \tilde a^0, \tilde a^1) &=& (\tilde u^1, \tilde a^0) \label{pi10}
\end{eqnarray}
from $T\pc^{1,3}$ to the $\tilde u^0-\tilde a^1$ and $\tilde u^1-
\tilde a^0$ planes,
respectively (see figure \ref{fig_msrspectrum}). From (\ref{a0is}) and (\ref{a1is}), we see that an orbit
corresponding to a spacetime trajectory of constant Minkowski curvature
$g$ projects under $\pi_{01}$ and $\pi_{10}$ to straight lines through
the origin of slope ${\mathfrak{a} g^{-1}}$ in the ${\tilde u^0-\tilde
a^1}$ and ${\tilde u^1-\tilde a^0}$ planes.
Special relativity allows arbitrarily high accelerations, hence \textsl{there} the
spectrum of uniformly accelerated spacetime curves is given by \textsl{all}
straight lines through the origin in these planes, with the same slope
in both planes for one particular curve.\\
In the framework presented here, however, $g$ is bounded from above by $\mathfrak{a}$. 

Using the conversion formula (\ref{tilde}), we see
that in the spacetime projection, (\ref{a0is}-\ref{a1is}) gives the
familiar hyperbolae of Minkowski curvature $g$, but only for
$g<\mathfrak{a}$. This determines
the spectrum of uniformly accelerated curves in extended special
relativity up to Poincare transformations (see figure \ref{fig_msrspectrum}).

%\begin{figure}
% \epsfig{file=msr_trafoscheme.eps,height=4cm, width=8cm}\\
% \includegraphics[-20,0][100,180]{msr_trafoscheme.eps}
% \caption{\label{fig_trafoscheme} transformation induced on 
% trajectories and orbits}
%\end{figure}

\subsection{\label{sec_trafogroup}Special Transformations in $SO_\pc(1,3)$}
Exponentiation of the generators $M^{\mu\nu}$ (cf. section \ref{sec_PLorentz}) yields
the familiar real Lorentz transformations acting on $T\pc^{1,3}$.
The group elements generated by the $I M^{\mu\nu}$ are given by the
ordinary Lorentz transformations evaluated with purely pseudo-imaginary
arguments. For notational simplicity,
we exhibit their properties in a $(1+1)$-dimensional theory. The
pseudo-boost of accelerity $\beta$ is then given by
\begin{equation} \Lambda_{\textrm{boost}}(I \beta) = 
  \left(\begin{array}{cc} 
         \,\, \cosh(\beta) & I \sinh(\beta)\\
         I \sinh(\beta) & \,\,\cosh(\beta)
      \end{array}\right),
\end{equation}
and its action on the eight-velocity $U = \tilde u + I \tilde a$ is
\begin{equation}
  \left(\cfvec{\tilde u^0}{\tilde u^1}{\tilde a^0}{\tilde a^1}\right) \mapsto 
  \left(\cfvec{\cosh(\beta)\,\tilde u^0 + \sinh(\beta)\,\tilde a^1}
    {\cosh(\beta)\,\tilde u^1 + \sinh(\beta)\,\tilde a^0}
    {\cosh(\beta)\,\tilde a^0 + \sinh(\beta)\,\tilde u^1}
    {\cosh(\beta)\,\tilde a^1 + \sinh(\beta)\,\tilde u^0}\right).
\end{equation}
This \textsl{hyperbolically rotates} a straight line of hyperbolic angle $\beta$ in the $\tilde
u^0-\tilde a^1$ and $\tilde u^1-\tilde a^0$ plane to another
straight line of hyperbolic angle $\alpha+\beta$ in the respective
plane. One can check that the the two remaining pseudo-boosts do exactly the
same with respect to the other space directions. Hence the
pseudo-boosts acting on the eight-velocity map the \textsl{projections} (\ref{pi01}-\ref{pi10}) of a $U$-curve to the
\textsl{projection} of another uniformly accelerated curve $U'$. We still
have to check whether it maps the \textsl{whole} curve U to a uniformly
accelerated curve $U'$. But this is easily seen by counting degrees of
freedom:
for a pseudo-boost in 1-direction, the projections of the transformed curve give us the two constraints
\begin{eqnarray}
  \tilde {a'}^0 &=& \tanh(\alpha+\beta) \tilde {u'}^1\\
  \tilde {a'}^1 &=& \tanh(\alpha+\beta) \tilde {u'}^0
\end{eqnarray} 
and we know that
\begin{eqnarray}
  \tilde {a'}^2 &=& a^2\\
  \tilde {a'}^3 &=& a^3\\
  \tilde {u'}^2 &=& u^2\\
  \tilde {u'}^3 &=& u^3
\end{eqnarray}
Further we know that any submaximally accelerated timelike curve lies
in the hypersurface
$$ \eta(U,U) = 1$$
in $U$-space. Being $O_\pc(1,3)$-transformations, pseudo-boosts respect this
condition, and thus it is still true for the transformed curve
$U'$. This leaves us with $8-(2+4+1) = 1$ degrees of freedom, uniquely
determining $U'$.\\ 
The pseudo-boosts transform the eight-velocity of a submaximally
accelerated orbit in one frame to the eight-velocity in a relatively accelerated frame.\\[5pt]
As $U=\frac{dX}{d\omega}$ and $d\omega$ is an $SO_\pc(1,3)$-invariant,
the pseudo-Lorentz group acts also linearly on the orbit $X$. The
action is non-linear, however, on the
spacetime curve $\pi(X)$. Note that the pseudo-Lorentz transformations
with not purely real parameter only induce a well-defined action on
the \textsl{space of motions} on spacetime, but cannot be understood
as a map $M \longrightarrow M$ of spacetime onto itself. This is so
because the components projected out by $\pi: TM \longrightarrow M$
mix with the spacetime coordinates under such transformations. Hence
spacetime coordinates fail to be well-defined under changes to
uniformly accelerated frames. Thus extended relativity anticipates the
Unruh \cite{qftMA} effect on a classical level already. This also
presents another manifestation of the non-commutative geometry on
spacetime induced by a finite upper bound on accelerations, as first
tentatively noted in (\ref{NCgeo}).\\
Pseudo-boosts in an arbitrary space direction can be
composed from the pseudo-boosts in the coordinate directions and an
appropriate rotation, exactly like for real boosts:
\begin{eqnarray}
  \textrm{rotation}^{-1}_R \quad \textrm{boost}_\mathbf{n} \quad
  \textrm{rotation}_R \quad &=& \quad \textrm{boost}_{R\mathbf{n}},\notag\\
  \textrm{rotation}^{-1}_R \quad \textrm{p-boost}_\mathbf{n} \quad
  \textrm{rotation}_R \quad &=& \quad \textrm{p-boost}_{R\mathbf{n}}.\notag
\end{eqnarray}
The role played by the \textsl{pseudo-rotations} is illuminated by the
identity
\begin{equation}
   \textrm{p-rot}^{-1}_i\left(\frac{\pi}{2}\right) \quad \textrm{boost}_i \quad
  \textrm{p-rot}_i\left(\frac{\pi}{2}\right) \quad = \quad
  \textrm{p-boost}_i.\notag
\end{equation}
%\begin{figure}[h]
%\includegraphics[-100,0][100,70]{msr_protations.eps}
%\caption{\label{fig_protations} relation between boost, props, rotations and protations}
%\end{figure}
The pseudo-rotations rotate velocities into accelerations and vice versa,
thus showing explicitly that there is no well-defined distinction
between velocities and accelerations, like in canonical classical
mechanics due to symplectic symmetry. We will see this mechanism at
work explicitly when discussing point particle dynamics in chapter \ref{sec_pointparticle}.

\subsection{Physical Postulates of Extended Special Relativity}
Equipped with the machinery developed above, we can now concisely
formulate the physical postulates of the maximal acceleration
extension of special relativity.\\
\textbf{Postulate I}.\\ Massive particles are described by
submaximally accelerated orbits $X$, i.e.
\begin{equation}\label{Uunit}
  \eta\left(dX,dX\right) > 0
\end{equation}
everywhere along $X$.\\
\textbf{Postulate II.} (modified clock postulate, \cite{Brandt2})\\ The physical time
measured by a clock with submaximally accelerated orbit $X$ is
given by the integral over the line element,
\begin{equation}
  \Omega = \mathfrak{a}^{-1} \int d\omega = \mathfrak{a}^{-1} \int \sqrt{\eta\left(\frac{dX}{d\lambda},
  \frac{dX}{d\lambda}\right)} \, d\lambda. 
\end{equation}

For curves of uniform accelerity $\alpha$, the modified clock
postulate gives a departure from the special relativity
prediction by a factor of 
\begin{equation}\label{MSRdeviation}
\cosh^{-1}(\alpha) = \sqrt{1-\frac{g^2}{\mathfrak{a}^2}}.
\end{equation} 

%\epsfig{file=msr_msrproper.eps,height=5cm,width=8cm}
%\caption{\label{fig_msrproper} proper time in SR and ESR dependent on
%acceleration $g$}

Hence experiments testing the clock postulate and involving
accelerations $|g|<g_{exp}$ give a lower bound on the hypothetical
maximal acceleration $\mathfrak{a}$.\\
Farley et al. \cite{Farley} have measured the decay rate of
muons with acceleration $g_{\textrm{exp}} = 5\times 10^{18} m s^{-2}$ 
within an
accuracy of $2$ percent. This corresponds to a measurement of the
lifetime within the same accuracy $\Delta=0.02$. Hence the factor
(\ref{MSRdeviation}) must deviate from unity less than $\Delta$: 
\begin{equation}
  \sqrt{1-\frac{g^2}{\mathfrak{a}^2}} > 1 - \Delta,
\end{equation} 
thus leading to an experimental lower bound for the maximal acceleration
\begin{equation}
   \mathfrak{a} > \frac{g}{\sqrt{\Delta\left(2-\Delta\right)}} \approx
   2.5\times 10^{19} m s^{-2}. 
\end{equation}
From the extended relativistic correction to the Thomas precession,
one obtains a much better upper bound $\mathfrak{a}\geq 10^{22} m
s^{-2}$, as shown in \cite{bikletter}.\\

\textbf{Correspondence Principle}\\  
From the postulates above, we recognize that in the limit $\mathfrak{a} \rightarrow\infty$, we have $d\omega \rightarrow d\tau$, and $U \rightarrow u$,
i.e. \textsl{extended special relativity becomes special relativity}.

\subsection{Born-Infeld Theory Revisited \label{sec_BIrevisited}}
After the formal developments in the last two chapters, we return to
the starting point of our investigations, the Born-Infeld
action. We demonstrate that the transformation of the commutation
relations (\ref{commrel}) as a second rank tensor of the
pseudo-complex Lorentz group is well-defined, and
hence Born-Infeld theory is compatible with the extended specially
relativistic kinematics developed earlier in this chapter. This shows
that the pseudo-complexification procedure was indeed successful in
the kinematization of the Born-Infeld symmetries associated with the
maximal acceleration. 

In section \ref{sec_dyngoeskin}, we assumed a particular coordinate
non-commutativity in order to recast the Born-Infeld action in the
suggestive form (\ref{BIcommform}). Now we are in a position to prove
that this is is the \textsl{only well-defined} non-commutative geometry
admitted by $SO_\pc(1,3)$ symmetry.  Assume the commutation relations
in the background of an electromagnetic field are more generally given
by
\begin{eqnarray}
  \left[x^\mu, x^\nu\right] &=& - i e K^{\mu\nu},\\
  \left[x^\mu, p^\nu\right] &=& - i \eta^{\mu\nu},\\
  \left[p^\mu, p^\nu\right] &=& i e F^{\mu\nu},
\end{eqnarray}
with an antisymmetric, otherwise (so far) arbitrary
$K^{\mu\nu}$. Using the notation ${X^m = \left(\mathfrak{a} x^\mu, u^\mu\right)}$, we define the tensor
$\tilde H$ through
\begin{equation}
  \left[X^m,X^n\right] = - i \frac{\mathfrak{a}}{m} \underbrace{\left(
   \begin{array}{cc}
      - e \mathfrak{a} m K^{\mu\nu} & - \eta^{\mu\nu} \\
       \eta^{\mu\nu} & - \frac{e}{a m} F^{\mu\nu}
   \end{array} 
  \right)}_{\equiv \tilde H^{mn}}
\end{equation}
and require it transforms as a second rank tensor under $SO_\pc(1,3)$:
\begin{equation}\label{Htransform}
  H^{mn} \longrightarrow {S^m}_a H^{ab} {S^n}_b = {S^m}_a H^{ab}
  {\left(S^t\right)_b}^n.
\end{equation}
The transformation rule (\ref{Htransform}) is \textsl{consistent} with what we expect
from the action of \textsl{real} Lorentz-transformations on $F$, $K$ and $g$:
\begin{equation}
\left(\begin{array}{cc}\hat K& g\\-g&\hat F\end{array}\right)
\longrightarrow
\left(\begin{array}{cc}\Lambda\hat K \Lambda^t& \Lambda g \Lambda^t
\\ - \Lambda g \Lambda^t & \Lambda\hat F \Lambda^t \end{array}\right).
\end{equation}
Now we calculate the action on $\tilde H$ of a pseudo-boost with accelerity $\beta$ in spatial $1$-direction. Consider the special case of
$F=K=0$ and $g=\eta$, i.e.
\begin{equation}\label{trivialH}
\tilde H^{ab} = \left(\begin{array}{cc}0 & {\eta^\alpha}_{\bar\beta} \\ - {\eta_{\bar\alpha}}^\beta & 0\end{array}\right).
\end{equation}
For calculational simplicity, we only consider the $(1+1)$-dimensional
case and we get from (\ref{trivialH}) using the transformation law (\ref{Htransform}):
\begin{equation}\label{H00trafo}
\tilde H^{ab} \longrightarrow \left(\begin{array}{cc|cc}
  0 & 2\,sh\,ch & sh^2+ch^2 & 0\\
  -2\,sh\,ch & 0 & 0 & -(sh^2+ch^2)\\[5pt]
\hline
  -(sh^2+ch^2) & 0 & 0 & -2\,sh\,ch\\
  0 & sh^2+ch^2& 2\,sh\,ch & 0
\end{array}\right),
\end{equation}
where $ch=\cosh(\beta)$, $sh=\sinh(\beta)$ and $\beta>0$, i.e. $S$ is a prop in the
\textsl{negative (!)} $x_1$ direction.\\
We observe two related points
\begin{enumerate}
 \item the $SO_\pc(1,3)$-transformation \textsl{preserves the antisymmetry} of $H$, which is of
 course crucial for the interpretation of the component blocks $K$,
 $F$ and $g$, and in turn for the well-definition of $H$.
 \item the mixing behaviour shows
 that, for reasons of consistency, it is necessary to identify the
 mixed parts of $K$ and $F$ up to constant factors:
 \begin{equation}
   K^{0i} = \frac{-1}{{m_0}^2 \mathfrak{a}^2} F^{0i}\qquad i=1,2,3\, .
 \end{equation}
\end{enumerate}
Hence we recognise that either the presence of an electromagnetic
field or the change to a relatively accelerated frame introduces a
($\mathfrak{a}^{-2}$-suppressed) time-position non-commutativity,
of \textsl{exactly} the form required to give (\ref{BIcommform}).\\
It is easy to see that the expression
\begin{equation}
 \det\left(H\right)
\end{equation} 
is invariant under the $SO_\pc(1,3)$-transformation (\ref{Htransform}).
Hence the Born-Infeld Lagrangian 
\begin{equation}
  \mathcal{L}_{\textrm{BI}} = \sqrt{\det\left(g_{\mu\nu} + b F_{\mu\nu}\right)}
\end{equation}
can be written as the manifestly $SO_\pc(1,3)$-invariant expression
\begin{equation}
  \mathcal{L}_{\textrm{BI}} =  \det\left(H\right)^{\frac{1}{4}} 
\end{equation}
for a \textsl{distinguished choice} of the Born-Infeld parameter, i.e. 
$b=|e|m^{-1}\mathfrak{a}^{-1}$, so that relation (\ref{maxBIacc}) now
\textsl{follows} from pseudo-complex Lorentz invariance!\\[10pt]

\section{Pseudo-Complex Manifolds}
Our findings in the flat case $\reals^{1,3}$ motivate a generalization
to a generally curved $n$-dimensional manifold $M$. In particular, the
isomorphism (\ref{isomorphy})
\begin{equation}
  \left(\pc^n, \eta\right) \cong \left(T\reals^n, \eta^D, \eta^H\right),
\end{equation}
suggests to consider the tangent bundle $TM$, equipped with two
metrics of signature $(2,2n-2)$ and $(n,n)$, respectively. Remarkably,
much of such a mathematical framework has already been developed by
Yano and others \cite{Yano1973} from a pure mathematical point of view,
which we can now give a physical interpretation from our insights
gained in the flat case.

\subsection{Lifts to the Tangent Bundle}
Several kinds of lifts of geometrical objects from a base manifold
$M$ to its tangent bundle $TM$ are introduced, and some of their
properties essential for our purposes are explored. We use local coordinates
for all our definitions, but everything can be made coordinate-free as
shown in e.g. \cite{Yano1973}.\\  

Throughout this chapter, $M$ denotes a differentiable
manifold, $g$ a metric and $\nabla$ a linear connection on $M$. $\pi:
TM \longrightarrow M$ denotes the canonical bundle projection. Let
$\{x^\mu\}$ be local coordinates on $M$. The \textsl{induced}
coordinates for a point $(x,y=y^\mu \partial_\mu)\in TM$ are $(x^\mu,y^\mu)$. 
The shorthand notations 
\begin{equation}
  \partial_\mu \equiv \frac{\partial}{\partial x^\mu} \qquad
  \partial_{\bar \mu} \equiv \frac{\partial}{\partial y^\mu} \qquad
  dx^{\bar \mu} \equiv dy^\mu
\end{equation}
are useful to clear up notation.\\ 
We now define the action of \textsl{vertical} and \textsl{horizontal}
lifts of functions, vectors and one-forms, and then algebraically
extend these definitions to tensors of arbitrary type \cite{Yano1973}.

\paragraph{Vertical Lifts} are defined on any differentiable manifold.
\begin{enumerate} 
  \item $f^V \equiv f \circ \pi$
  \item $\left(\partial_\mu\right)^V \equiv \partial_{\bar \mu}$
  \item $\left(dx^\mu\right)^V \equiv dx^\mu$
  \item algebraic extension via
           $$\left(P + Q\right)^V = P^V + Q^V$$
           $$\left(P \otimes Q\right)^V = P^V \otimes Q^V$$ 
\end{enumerate} 

\paragraph{Horizontal Lifts} are defined on manifolds carrying a
linear connection $\nabla$ with Christoffel symbols ${{\Gamma_\sigma}^\mu}_\tau$.
\begin{enumerate}
  \item $f^H \equiv 0$
  \item $\left(\partial_\mu\right)^H \equiv \partial_\mu - {\Gamma^\tau}_\mu
  \partial_{\bar \tau}$
  \item $\left(dx^\mu\right)^H \equiv dx^{\bar \mu} + {\Gamma^\mu}_\tau dx^\tau$
  \item algebraic extension via
           $$\left(P + Q\right)^H = P^H + Q^H$$
           $$\left(P \otimes Q\right)^H = P^H \otimes Q^V + P^V
            \otimes Q^H$$ 
\end{enumerate}
where ${\Gamma^\mu}_\tau \equiv y^\sigma {{\Gamma_\sigma}^\mu}_\tau$.\\
A third type of lifts to the tangent bundle, \textsl{diagonal} lifts,
are only defined for $(0,2)$ tensors on a manifold with symmetric
affine connection:
\paragraph{Diagonal Lift} Let $G$ be a $(0,2)$ tensor on $\left(M,\Gamma\right)$
\begin{equation}
G^D \equiv \left(G_{\mu\nu}\right)^V \left(dx^\mu\right)^V \otimes
\left(dx^\nu\right)^V + \left(G_{\mu\nu}\right)^V \left(dx^\nu\right)^H
\otimes \left(dx^\mu\right)^H
\end{equation}
As in the flat case, the \textsl{natural lift} $x^*$ of a curve $x: \reals
\longrightarrow M$ plays an important r\^ole. It is given in induced
coordinates by
\begin{eqnarray}
  X: \reals &\longrightarrow& TM\notag\\
    X^m &\equiv& \left(\mathfrak{a} x^\mu, \frac{dx^\mu}{d\tau}\right), 
\end{eqnarray}
where $d\tau$ is the arc length of the curve $x$ on $(M,g)$.

\subsection{Adapted Frames}
The induced frame on $TM$,
\begin{equation}
  \partial_\mu, \qquad \partial_{\bar\mu}
\end{equation}
allows an easy interpretation of a tangent bundle vector, but for many calculational purposes, the so-called
\textsl{adapted frame} is more convenient. The $2n$ local vector
fields 
\begin{equation}
  e_{(\mu)} \equiv \left(\partial_\mu\right)^H, \qquad
  e_{(\bar\mu)} \equiv \left(\partial_\mu\right)^V
\end{equation}
constitute a basis of the tangent space $T_X(TM)$ of the tangent bundle
at point $X\in TM$, the so-called adapted frame. The components of
these basis vectors in induced coordinates are
\begin{equation}
  e_{(\mu)} = \left(\begin{array}{c} \delta_\mu^\nu \\ -{\Gamma_\mu}^\nu
  \end{array}\right), \qquad e_{(\bar \mu)} = \left(\begin{array}{c}
  0 \\ \delta_\mu^\nu \end{array}\right).
\end{equation}
The dual coframe is 
\begin{equation}
  \epsilon^{\mu} \equiv \left(dx^\mu\right)^H, \qquad \epsilon^{\bar\mu} \equiv \left(dx^\mu\right)^V.
\end{equation}
In the adapted frame, horizontal and vertical lifts are similarly
easy to perform. Consider the vector $X=X^\mu \partial_\mu$ and the
one-form $\omega = \omega_\mu
dx^\mu$. In the adapted frame,
\begin{eqnarray}
  X^H = \left(\begin{array}{c} X^\mu \\ 0 \end{array}\right), &\qquad&
  X^V = \left(\begin{array}{c} 0 \\ X^\mu \end{array}\right) \\
 \omega^H = (0, \omega_\mu), \quad&\qquad& \omega^V = (\omega_\mu, 0). 
\end{eqnarray}
The transformation between induced and adapted coordinates has unit
determinant, as follows from \cite{Yano1973}.

\subsection{Lifts of the Spacetime Metric}

The above definitions allow to calculate the component form of the
horizontal and diagonal lifts of the metric $g$ of a semi-Riemannian
manifold $\left(M,g\right)$ with the connection being the
Levi-Civit\`a connection.\\
In induced coordinates,
\begin{eqnarray}
  g^D &=& \left(\begin{array}{cc} g_{ij} + g_{ts}{\Gamma^t}_i{\Gamma^s}_j
  & {\Gamma^t}_j g_{ti} \\ {\Gamma^t}_i g_{tj} & g_{ij} \end{array}\right),\\
  g^H &=& \left(\begin{array}{cc} \partial g_{ij} & g_{ij} \\ g_{ji} & 0 \end{array}\right),
\end{eqnarray}   
where $\partial \equiv y^i \partial_i$.\\
Using the adapted frame, this becomes
\begin{eqnarray}
  g^D &=& \left(\begin{array}{cc} g_{\mu\nu} & 0  \\ 0 & g_{\mu\nu}
  \end{array}\right) = g_{\mu\nu} \otimes 1,\\
  g^H &=& \left(\begin{array}{cc} 0 & g_{\mu\nu} \\ g_{\mu\nu} & 0
  \end{array}\right) = g_{\mu\nu} \otimes I.
\end{eqnarray}
One can check that for a metric $g$ of signature $(p,q)$ on $M$, both
$g^D$ and $g^H$ are globally defined, non-degenerate metrics on $TM$
with signature $(2p,2q)$ and $(p+q,p+q)$, respectively. Observe that
in the flat case $g \equiv \eta$, we have
\begin{eqnarray}
  g^D &=& \eta \otimes 1,\\
  g^H &=& \eta \otimes I, 
\end{eqnarray}
also in the induced frame, which we used throughout chapter
\ref{sec_MSR}.

\subsection{Connections on the Tangent Bundle}
Let $\tilde\nabla$ be the (torsion free) Levi-Civit\`a connection on
$TM$ with respect to $g^H$, i.e.
\begin{equation}
  \tilde\nabla g^H = 0.
\end{equation} 
Denote the Christoffel symbols $\tilde\Gamma$. In terms of the
Christoffel symbols $\Gamma$ formed with the metric $g$ on $M$, we
find
\begin{eqnarray}
   \chris{\tilde\Gamma}{\mu}{\tau}{\nu} &=& \chris{\Gamma}{\mu}{\tau}{\nu}, \notag\\
  \chris{\tilde\Gamma}{\mu}{\bar \tau}{\nu} &=& \partial
             \chris{\Gamma}{\mu}{\tau}{\nu}, \notag\\
  \chris{\tilde\Gamma}{\mu}{\bar \tau}{\bar \nu} &=& \chris{\Gamma}{\mu}{\tau}{\nu},\\
  \chris{\tilde\Gamma}{\bar \mu}{\bar \tau}{\nu} &=& \chris{\Gamma}{\mu}{\tau}{\nu},\notag
\end{eqnarray}    
all unrelated symbols of $\tilde \Gamma$ being zero.\\
The serious drawback of this connection is that in general
\begin{equation}
  \tilde\nabla g^D \neq 0.
\end{equation}
At the expense of introducing torsion on the tangent bundle, however,
we can find a connection $\nabla^H$ which makes both metrics
simultaneously covariantly constant:
\begin{eqnarray}
  \nabla^H g^H &=& 0,\\
  \nabla^H g^D &=& 0.
\end{eqnarray}
Denoting its Christoffel symbols $\hat\Gamma$, they are in terms of
the Christoffel symbols $\Gamma$:
\begin{eqnarray}
  \chris{\hat\Gamma}{\mu}{\tau}{\nu} &=& \chris{\Gamma}{\mu}{\tau}{\nu},\notag \\
  \chris{\hat\Gamma}{\mu}{\bar \tau}{\nu} &=& \partial
             \chris{\Gamma}{\mu}{\tau}{\nu} - {R_{\sigma\mu\nu}}^\tau y^\sigma,\notag \\
  \chris{\hat\Gamma}{\mu}{\bar \tau}{\bar \nu} &=& \chris{\Gamma}{\mu}{\tau}{\nu},\\
  \chris{\hat\Gamma}{\bar \mu}{\bar \tau}{\nu} &=& \chris{\Gamma}{\mu}{\tau}{\nu},\notag
\end{eqnarray}
all unrelated components of $\hat\Gamma$ being zero, and $R$ is the
Riemann curvature of $\Gamma$.\\
The torsion on $TM$ is then
\begin{equation}
  2 \chris{\hat\Gamma}{[\nu}{\bar \tau}{\mu]} = {R_{\mu\nu\sigma}}^\tau y^\sigma \qquad \textrm{(all
  unrelated vanishing)}
\end{equation}
in induced coordinates.\\
The connection $\nabla^H$ has the further nice properties
\begin{enumerate}
  \item $\nabla^H$ has vanishing Riemann tensor iff $\nabla$ has
  vanishing Riemann tensor. \cite{YI6}
  \item  $\nabla^H$ has vanishing Ricci tensor iff $\nabla$ has
  vanishing Ricci tensor. (\cite{Yano1973}, IV.4.3)
  \item  $\nabla^H$ has $g^D$-Ricci scalar $R$ if $\nabla$ has Ricci
scalar $R$ (see section \ref{sec_fieldeqns}).
  \item  $\nabla^H$ has vanishing $g^H$-Ricci scalar. (\cite{Yano1973}, IV.4.4)
\end{enumerate}
We will make use of these curvature properties when lifting the
Einstein field equations in section \ref{sec_fieldeqns}.

\subsection{Orbidesics}
Exactly as in the flat case, a curve $X: \reals \longrightarrow TM$ is
called an \textsl{orbit}, if there exists a frame such that
\begin{equation}\label{orbitcon}
  X = \pi(X)^*,
\end{equation}
i.e. if the natural lift of its bundle projection recovers the
curve. Again, in induced coordinates $X^m = \left(x^\mu,
y^\mu\right)$, this is equivalent to $y^\mu = \frac{dx^\mu}{d\tau}$,
where $\tau$ is the arc length of the curve $x$ in $(M,g)$.\\
As we are dealing with real manifolds, the orbital condition
(\ref{orbitcon}) also ensures that the projection $\pi(X)$ is always
timelike. However, in the general curved case, the distinguished
r\^ole of the orbits will be seen of much wider importance. We
therefore introduce the notions of orbidesics and orbiparallels. An
orbit $X$ which is also an autoparallel with respect to some connection
$\nabla$ is called a \textsl{$\nabla$-orbiparallel}. If, more
specially, the connection is the Levi-Civit\`a connection of some
metric $G$, then $X$ is called a \textsl{$G$-orbidesic}.\\
The following statements follow from (\cite{Yano1973},I.9.1 and I.9.2)
\begin{enumerate}
  \item Let $X$ be a $g^H$-orbidesic on $TM$. Then $\pi(X)$ is a
  geodesic on $\left(M,g\right)$.
  \item Let $x$ be a geodesic on $\left(M,g\right)$. Then $x^*$ is a
  $g^H$-orbidesic \textsl{and} a $g^D$-orbidesic.
\end{enumerate}
A direct corollary is that any $g^H$-orbidesic is a $g^D$-orbidesic,
but the converse does not hold.\\
Diagrammatically,\\

%%%%%%% ske_dia1.latex, from ske_dia1.fig
\setlength{\unitlength}{3947sp}%
\begingroup\makeatletter\ifx\SetFigFont\undefined%
\gdef\SetFigFont#1#2#3#4#5{%
  \reset@font\fontsize{#1}{#2pt}%
  \fontfamily{#3}\fontseries{#4}\fontshape{#5}%
  \selectfont}%
\fi\endgroup%
\begin{picture}(4512,1489)(526,-1319)
\thinlines
\put(1876,-136){\vector( 1,-1){750}}
\put(2476,-961){\vector(-1, 1){750}}
\put(4276,-961){\vector( 1, 1){750}}
\put(2551, 89){\line( 1, 0){150}}
\put(2851, 89){\line( 1, 0){150}}
\put(3151, 89){\line( 1, 0){150}}
\put(3451, 89){\line( 1, 0){150}}
\put(3751, 89){\line( 1, 0){150}}
\put(4051, 89){\line( 1, 0){150}}
\put(4351, 89){\vector( 1, 0){225}}
\put(526, 14){\makebox(0,0)[lb]{\smash{\SetFigFont{12}{14.4}{\familydefault}{\mddefault}{\updefault}
$g^H$-orbidesic on TM
}}}
\put(4726, 14){\makebox(0,0)[lb]{\smash{\SetFigFont{12}{14.4}{\familydefault}{\mddefault}{\updefault}
$g^D$-orbidesic on TM
}}}
\put(2776,-1261){\makebox(0,0)[lb]{\smash{\SetFigFont{12}{14.4}{\familydefault}{\mddefault}{\updefault}
$g$-geodesic on M
}}}
\put(1951,-811){\makebox(0,0)[lb]{\smash{\SetFigFont{12}{14.4}{\familydefault}{\mddefault}{\updefault}
*
}}}
\put(4801,-811){\makebox(0,0)[lb]{\smash{\SetFigFont{12}{14.4}{\familydefault}{\mddefault}{\updefault}
*
}}}
\put(2326,-436){\makebox(0,0)[lb]{\smash{\SetFigFont{12}{14.4}{\familydefault}{\mddefault}{\updefault}
$\pi$
}}}
\end{picture}

%%%% END of ske_dia1.latex
\vspace{0.8cm}
The $g^D$-line element of an orbit $X$ is denoted $d\omega$ and given
by
\begin{equation}
  d\omega^2 = g^D\left(dX,dX\right).
\end{equation}   
The spacetime line element of a bundle 
projection of an orbit $X$ is given by
\begin{equation}
  d\tau^2 = g(d\pi(X), d\pi(X)).
\end{equation}
An important observation are the relations
\begin{eqnarray}
 g^H(u^H,u^H)&=&0, \qquad\qquad\qquad \textrm{(\cite{Yano1973},p. 140)}\\
 g^D(u^H,v^H)&=&\left[g(u,v)\right]^V, \qquad\, \textrm{(\cite{Yano1973},IV.5.1)}
\end{eqnarray}
for vector fields $u, v$ on $M$. These relations allow to recognize
that the unit tangent vector to a $g^H$-orbidesic $X$ is just the
horizontal lift of the four-velocity of $\pi(X)$,
\begin{equation}
  \frac{dX}{d\omega} = \left(\frac{dx}{d\tau}\right)^H.
\end{equation}
This can be seen as follows.
\begin{equation}
  \left(\frac{dx}{d\tau}\right)^H = \left(\frac{dx^\mu}{d\tau}, -
  \frac{dx^\alpha}{d\tau} {{\Gamma_\alpha}^\mu}_\beta
  \frac{dx^\beta}{d\tau}\right) = \frac{dX}{d\tau}
\end{equation}
using the geodesic property of $x=\pi(X)$. Hence,
\begin{equation}
  d\omega^2 = g^D\left(\frac{dX}{d\tau},\frac{dX}{d\tau}\right)\,d\tau^2 = g\left(\frac{dx}{d\tau}, \frac{dx}{d\tau}\right) d\tau^2 = d\tau^2   
\end{equation} 
using the second relation above. Thus, 
\begin{equation}\label{orbidesiclength}
  g^D\left(dX,dX\right) > 0 \qquad \textrm{for any } g^H-\textrm{orbidesic}.
\end{equation}
For orbits which are \textsl{not} $g^H$-orbidesical, this is neither
generally true nor generally false. This allows the following classification: An
orbit $X$ is called \textsl{submaximally accelerated}, if
$g^D\left(dX,dX\right) > 0$ everywhere along the orbit. The result (\ref{orbidesiclength}) is
reassuring, as orbidescial motion will represent
\textsl{unaccelerated} motion.

\subsection{Orbidesic Equivalence}
We found that there is a one-to-one correspondence between
$g$-geodesics and $g^H$-orbidesics. However, in the next section we
will explain why $\nabla^H$ rather than $\tilde\nabla$ is the
appropriate connection on $TM$, and it is desirable to learn about the
relation between $g$-geodesics and $\nabla^H$-orbiparallels. The
remarkable observation of this short section will be that the
$g^H$-orbidesics are the $\nabla^H$-orbiparallels, and vice versa.\\
The following statements are shown in (\cite{Yano1973}, II.9.1 and II.9.2).
\begin{enumerate}
  \item Let $X$ be an $\nabla^H$-orbiparallel. Then $\pi(X)$ is a $g$-geodesic.
  \item Let $x$ be a $g$-geodesic. Then $x^*$ is a $\nabla^H$-orbiparallel.
\end{enumerate}
This completes the diagram from the last section to \\

%%%%%%% ske_dia2.latex, from ske_dia2.fig

\setlength{\unitlength}{3947sp}%
\begingroup\makeatletter\ifx\SetFigFont\undefined%
\gdef\SetFigFont#1#2#3#4#5{%
  \reset@font\fontsize{#1}{#2pt}%
  \fontfamily{#3}\fontseries{#4}\fontshape{#5}%
  \selectfont}%
\fi\endgroup%
\begin{picture}(4512,2614)(526,-2444)
\thinlines
\put(1876,-136){\vector( 1,-1){750}}
\put(2476,-961){\vector(-1, 1){750}}
\put(4276,-961){\vector( 1, 1){750}}
\put(1726,-2011){\vector( 1, 1){750}}
\put(2626,-1336){\vector(-1,-1){750}}
\put(2551, 89){\line( 1, 0){150}}
\put(2851, 89){\line( 1, 0){150}}
\put(3151, 89){\line( 1, 0){150}}
\put(3451, 89){\line( 1, 0){150}}
\put(3751, 89){\line( 1, 0){150}}
\put(4051, 89){\line( 1, 0){150}}
\put(1126,-436){\line( 0,-1){150}}
\put(1126,-736){\line( 0,-1){150}}
\put(1126,-1636){\line( 0,-1){150}}
\put(1126,-286){\vector( 0, 1){225}}
\put(1126,-1036){\line( 0,-1){150}}
\put(1126,-1336){\line( 0,-1){150}}
\put(1126,-1936){\vector( 0,-1){225}}
\put(4351, 89){\vector( 1, 0){225}}
\put(526, 14){\makebox(0,0)[lb]{\smash{\SetFigFont{12}{14.4}{\familydefault}{\mddefault}{\updefault}
$g^H$-orbidesic on TM
}}}
\put(4726, 14){\makebox(0,0)[lb]{\smash{\SetFigFont{12}{14.4}{\familydefault}{\mddefault}{\updefault}
$g^D$-orbidesic on TM
}}}
\put(2776,-1261){\makebox(0,0)[lb]{\smash{\SetFigFont{12}{14.4}{\familydefault}{\mddefault}{\updefault}
$g$-geodesic on M
}}}
\put(1951,-811){\makebox(0,0)[lb]{\smash{\SetFigFont{12}{14.4}{\familydefault}{\mddefault}{\updefault}
*
}}}
\put(4801,-811){\makebox(0,0)[lb]{\smash{\SetFigFont{12}{14.4}{\familydefault}{\mddefault}{\updefault}
*
}}}
\put(2326,-436){\makebox(0,0)[lb]{\smash{\SetFigFont{12}{14.4}{\familydefault}{\mddefault}{\updefault}
$\pi$
}}}
\put(526,-2386){\makebox(0,0)[lb]{\smash{\SetFigFont{12}{14.4}{\familydefault}{\mddefault}{\updefault}
$\nabla^H$-orbiparallel on TM
}}}
\put(2401,-1861){\makebox(0,0)[lb]{\smash{\SetFigFont{12}{14.4}{\familydefault}{\mddefault}{\updefault}
*
}}}
\put(1801,-1561){\makebox(0,0)[lb]{\smash{\SetFigFont{12}{14.4}{\familydefault}{\mddefault}{\updefault}
$\pi$
}}}
\end{picture}

%%%%%% END of ske_dia2.latex

\subsection{The Tachibana-Okumura No-Go Theorem\label{sec_nogo}}
In 1981, Caianiello observed that requiring positivity of tangent
bundle curves with respect to the metric
\begin{equation}
  \left(\begin{array}{cc} \eta & 0 \\ 0 & \eta \end{array}\right),
\end{equation}
which generalizes to $g^D$ in the curved case, introduces an upper
bound on worldline accelerations \cite{Caianiello1981}. 
Given the importance of the complex structure of the phase space in
Hamiltonian mechanics, it seemed quite sensible to establish a complex
structure also on the tangent or cotangent bundle of Minkowski spacetime, and several
attempts have been made in both the flat \cite{Low} and the curved
case \cite{Brandt}.\\

We are now in a position to prove that a complex structure $F$ on the
tangent bundle is \textsl{incompatible} with the assumption of a
maximal acceleration introduced by $g^D$ in the sense explained
above. This statement is true under the physical assumption that a
strong principle of equivalence between the flat and the curved case
holds, or in mathematical terms, that the structures $g^D$ and $F$ are
required to be simultaneously covariantly constant:
\begin{eqnarray}
  \bar \nabla g^D &\stackrel{!}{=}& 0,\\
  \bar \nabla F\,\,\,   &\stackrel{!}{=}& 0.
\end{eqnarray}
In this approach, $g^D$ is the only metric on $TM$, and so we take $\bar
\nabla$ to be the Levi-Civit\`a connection with respect to $g^D$. The
globally defined one-form on $TM$
\begin{equation}
  \Theta \equiv - g_{\mu\nu} y^\mu dx^\nu
\end{equation} 
has an exterior differential
\begin{equation}
  d\Theta = \frac{1}{2} F_{mn} dX^m dX^n.
\end{equation}
Raising and lowering indices with $g^D$, we can easily verify that
\begin{equation}
  {F_a}^m {F_n}^a = - \delta^m_n,
\end{equation}
hence ${F_m}^n$ defines an \textsl{almost complex structure} on
$TM$. The covariant derivative of ${F_m}^n$ with respect to
$\bar\nabla$ evaluates to \cite{Yano1973} 
\begin{eqnarray}
  \bar\nabla_\mu {F_\nu}^\alpha = - \bar\nabla_\mu
  {F_{\bar\nu}}^{\bar\alpha} &=& \frac{1}{2}
  \left({R^\alpha}_{\mu\sigma\nu} +
  {R_{\mu\nu\sigma}}^{\alpha}\right)y^\sigma,\\
  \nabla_{\bar\mu} {F_\nu}^\alpha = \nabla_{\bar\mu} {F_\nu}^{\alpha}
  &=& \frac{1}{2} {R^\alpha}_{\nu\sigma\mu} y^\sigma,
\end{eqnarray} 
and all other terms vanish. This immediately gives the

\paragraph{Tachibana-Okumura No-Go Theorem}  \cite{TO1962}\\
The tangent bundle $TM$ of a semi-Riemannian manifold $M$ has
\textsl{simultaneously covariantly constant} metric $g^D$ and complex structure
$F$ if and only if the \textsl{base manifold $M$ is flat}.\\

This makes all past approaches in this direction physically
questionable. As pointed out in section \ref{sec_dyngoeskin}, even
though for \textsl{flat} Minkowski spacetime, equipping the tangent bundle with
a complex structure \textsl{is} compatible with a maximal
acceleration, the slightest perturbation would render the
theory inconsistent, and a generalization to the curved case is
entirely frustrated.\\

Clearly, the \textsl{pseudo-complex} approach presented in this paper
circumvents the Tachi\-bana-Okumura no-go theorem. 

\section{Maximal Acceleration Extension of General Relativity}
The stage of extended general relativity is the tangent
bundle $TM$ of curved spacetime $M$, equipped with the horizontal and
diagonal lift of the spacetime metric $g$, $\left(TM, g^H,
g^D\right)$.
As the linear connection we take $\nabla^H$. Then both metrics are
covariantly constant:
\begin{eqnarray}
  \nabla^H \, g^H &=& 0,\\
  \nabla^H \, g^D &=& 0,
\end{eqnarray}
and we know that $\nabla^H$ has the same orbidesics as
$\tilde\nabla$, the Levi-Civit\`a connection of $g^H$. 
Hence we can establish the strong principle of equivalence within this
framework, circumventing the Tachibana-Okumura no-go-theorem.

\subsection{Physical Postulates of Extended General Relativity\label{sec_MGRpostulates}}

We require for \textsl{all orbits} representing submaximally
accelerated particle motion that 
\begin{enumerate}
\item $g^H(dX,dX) = 0$,
\item $g^D(dX,dX) > 0$.
\end{enumerate}
For orbidesics, we saw these conditions are automatically satisfied.\\

We measure physical time along a submaximally accelerated orbit $X$ by
$$d\omega = \left[g^D(dX,dX)\right]^{-\frac{1}{2}}.$$

Particles, under the influence of gravity only, travel along
$g^H$ (null) orbidesics, or equivalently, $\nabla^H$-orbiparallels.\\   

Hence for particles in a gravitational field only,
the orbits are also $g^D$-geodesics, and their proper time is measured 
with $g^D$ as well. One could say that we only really \textsl{need} the \textsl{two}
metrics $g^H$ and $g^D$ if it comes to non-geodesic orbits. This is an
explanation, why in general relativity on
spacetime with small (non-gravitational) acceleration, one metric
always seemed to be enough.\\
In the absence of any force besides gravity, extended general
relativity is exactly equivalent to general relativity, as desired. 

\subsection{Field Equations \label{sec_fieldeqns}}
In order to achieve a formulation of extended general relativity
entirely in terms of tangent bundle concepts, it is certainly
necessary to lift the field equations for $g$ on spacetime to field
equations for $g^H$ and $g^D$ on the tangent bundle.\\
The Ricci tensor $\hat R$ of the connection $\nabla^H$ evaluates to 
\begin{equation}
  \hat R_{ab} = \left(\begin{array}{cc} R_{\alpha\beta} & 0 \\ 0 & 0 \end{array}\right),
\end{equation}
in induced \textsl{and} adapted coordinates alike. We recognize that
\begin{equation}
  \hat R_{ab} = \left(R_{\alpha\beta}\right)^V.
\end{equation}
Unlike the Ricci tensor, the Ricci scalar depends on the metric used
for its contraction,
\begin{eqnarray}
  \hat R_{ab} \left(g^H\right)^{ab} = 0,\\
  \hat R_{ab} \left(g^D\right)^{ab} = R,
\end{eqnarray}
where $R\equiv R_{\alpha\beta}g^{\alpha\beta}$ is the curvature scalar on
$M$. This can be immediately seen in the adapted frame. Thus it is
sensible to define the Ricci scalar $\hat R$ on $TM$ as
\begin{equation}
  \hat R \equiv  \hat R_{ab} \left(g^D\right)^{ab}.
\end{equation}
Now consider the Einstein field equations on $M$,
\begin{equation}\label{EFE}
  R^{\alpha\beta} - \frac{1}{2} g^{\alpha\beta} R = 8 \pi G T^{\alpha\beta}.
\end{equation}
'Duplication' trivially gives an equivalent set of equations
\begin{equation}\label{doubleEFE}
  \left(\begin{array}{cc}R^{\alpha\beta}&0\\0&R^{\alpha\beta}\end{array}\right)
- \frac{1}{2}
\left(\begin{array}{cc}g^{\alpha\beta}&0\\0&g^{\alpha\beta}\end{array}\right)
R = \left(\begin{array}{cc}T^{\alpha\beta}&0\\0&T^{\alpha\beta}\end{array}\right)
\end{equation}
Observe that the first term on the left can be interpreted in the
adapted frame as
\begin{equation}
\left[\left(g^D\right)^{am}\left(g^D\right)^{bn} +
\left(g^H\right)^{am}\left(g^H\right)^{bn} \right] \hat R_{mn}.
\end{equation}
The second term equals
\begin{equation}
  - \frac{1}{2} \left[\left(g^D\right)^{ab}\left(g^D\right)^{mn} +
  \underbrace{\left(g^H\right)^{ab}\left(g^H\right)^{mn}}_{= 0}\right] \hat R_{mn}.
\end{equation}
Defining the 'double metric'
\begin{equation}
  G^{abcd} \equiv \left(g^D\right)^{ab}\left(g^D\right)^{cd} +
\left(g^H\right)^{ab}\left(g^H\right)^{cd},
\end{equation}
we find from (\ref{doubleEFE}) the tangent bundle tensor equation
\begin{equation}\label{liftedEFE}
  \left(G^{ambn} - \frac{1}{2} G^{abmn}\right) \hat R_{mn} = 8 \pi G\, \hat T^{ab},
\end{equation}
where 
\begin{equation}
  \hat T^{ab} \equiv \left(T^{\alpha\beta}\right)^D.
\end{equation}
The equations (\ref{liftedEFE}) are fully equivalent to the Einstein
field equations (\ref{EFE}), and we call them the \textsl{lifted field
equations}. Being a tensor equation, (\ref{liftedEFE}) is valid in any
frame, not just the adapted frame we used for its derivation. 

\section{Point Particle Dynamics}\label{sec_pointparticle}

\subsection{Free Massive Particles}
In a series of carefully written papers, Nesterenko et al. \cite{Nesterenko,Nestsecond,Nestaction}
investigate into the Lorentz invariant action
\begin{equation}\label{Nestaction}
  S = \int \sqrt{\mathfrak{a}^2 - g^2} d\tau, \qquad \textrm{where } d\tau^2=dx^\mu
  dx^\mu, \quad g^2 = - \frac{d^2x^\mu}{d\tau^2} \frac{d^2x_\mu}{d\tau^2}
\end{equation}
within the framework of special relativity. As the associated
Lagrangian depends on first and second order derivatives,
\begin{equation}
  L = L\left(\dot x^\mu, \ddot x^\mu\right),\qquad \dot{} \equiv \frac{d}{dt}, 
\end{equation}
the Euler-Lagrange equations are
\begin{equation}
  \frac{d^2}{dt^2} \frac{\partial L}{\partial \ddot x^\mu} -
  \frac{d}{dt} \frac{\partial L}{\partial \dot x^\mu} + \frac{\partial
  L}{\partial x^\mu} = 0,
\end{equation}
and transition to the Hamiltonian formalism is more complicated as for
first order Lagrangians, though possible. It is shown in
\cite{Nestsecond, Nestaction}
that the action (\ref{Nestaction}) provides viable specially
relativistic dynamics for a free massive particle, consistent with the
assumption of an upper bound $\mathfrak{a}$ on accelerations.\\
Nevertheless, the appearance of Lagrangians with second order
derivatives is a considerable inconvenience, and makes the theory look
prone to encountering difficulties at a later stage,
e.g. quantization, even if it can be shown to be consistent in
particular cases as above.\\
However, the second order derivatives in (\ref{Nestaction}) were
introduced in order to \textsl{dynamically enforce} a maximal
acceleration in the framework of special relativity. Writing the
action in the manifestly $O_\pc(1,3)$-invariant form
\begin{equation}\label{MSRaction}
  S = \int d\omega = \int \sqrt{\dot X^\mu \dot X^\mu} dt,\qquad X^\mu
  \equiv x^\mu + I u^\mu,
\end{equation}
and, sloppily speaking, leaving the rest to the extended relativistic
\textsl{kinematics}, 'miraculously' solves the problems mentioned
above: the relation between the four-velocity and four-acceleration is
absorbed in the pseudo-complex tangent bundle geometry, and hence
(\ref{MSRaction}) is \textsl{not} just a notational trick. 
Moreover, the pseudo-complexification prescription
\begin{equation}\label{pcp}
   \reals^{1,3} \longrightarrow \pc^{1,3}
\end{equation}
turns out to be equally applicable to Lagrangians, in the present case
converting the action of a free relativistic point particle to the extended
relativistic action (\ref{MSRaction}), automatically generating the necessary
constraints!\\
Start with a specially relativistic point particle of mass $m$,
\begin{equation}
  S = m \int \sqrt{\dot x^\mu \dot x^\mu} dt.
\end{equation}  
It is convenient to rewrite the Lagrangian in 'Hamiltonian form', explicitly
including the constraint on the associated canonical momenta:
\begin{equation}
  L = p_\mu \dot x^\mu - \frac{1}{2} \lambda \left(p_\mu p^\mu - m^2\right).
\end{equation}
We apply the pseudo-complexification prescription (\ref{pcp}) and
replace 
\begin{eqnarray}
  x^\mu \longrightarrow X^\mu &\equiv& x^\mu + I u^\mu,\\
  p^\mu \longrightarrow P^\mu &\equiv& p^\mu + I f^\mu,
\end{eqnarray}
obtaining 
\begin{equation}\label{LMSR}
  L_{\textrm{MSR}} = P_\mu \dot X^\mu - \frac{1}{2}
  \lambda\left(P_\mu P^\mu - m^2\right).
\end{equation}
Note that naively performing derivatives
\begin{equation}
  \frac{\partial}{\partial X^\mu} \qquad \textrm \qquad
  \frac{\partial}{\partial \dot X^\mu}
\end{equation}
will lead into trouble, as $\pc$ is only a ring, and hence the
differential quotient is not generally defined. However, from the
fully equivalent \textsl{tangent bundle} point of view, the definition
\begin{equation}
   \frac{\partial}{\partial X^\mu} \equiv \frac{1}{2}
   \left(\frac{\partial}{\partial x^\mu} + I \frac{\partial}{\partial u^\mu} \right)
\end{equation}
is easily understood, and turns out to be a useful one. Thus we see
that
\begin{equation}
  \frac{\partial L}{\partial \dot X^\mu} = p_\mu + I f^\mu = P_\mu
\end{equation}
really gives $P_\mu$ as the canonical momentum conjugate to $X^\mu$,
as suggested by the notation.\\
We obtain the equations of motion from (\ref{LMSR}) by variation with
respect to $e$, $P$, and $X$:
\begin{eqnarray}
  P_\mu P^\mu &=& m^2, \label{eq_i}\\
  \dot X^\mu - \lambda P^\mu &=& 0, \label{eq_ii}\\
  \frac{d}{dt} P^\mu &=& 0 \label{eq_iii}.
\end{eqnarray}
Relations (\ref{eq_ii}) and (\ref{eq_iii}) immediately give
\begin{equation}\label{straight}
  \ddot X^\mu = 0,
\end{equation}
and from (\ref{eq_i}) and (\ref{eq_ii})
\begin{equation}
  \dot X^\mu \dot X_\mu = \lambda^2 P^\mu P_\mu = \lambda^2 m^2 > 0, 
\end{equation}
hence the particle is submaximally accelerated! We choose the gauge
$\lambda=m^{-1}$, corresponding to natural parameterization $\omega$,
with $d\omega^2=dX^\mu dX_\mu$. Then from (\ref{straight}),
\begin{equation}
  \dot X^\mu = c^\mu + I d^\mu,
\end{equation}
for constant real four-vectors $c, d \in \reals^{1,3}$ satisfying
\begin{eqnarray}
  c^\mu c_\mu + d^\mu d_\mu &=& 1\label{eq_a}\\
  c^\mu d_\mu &=& 0 \label{eq_b}.
\end{eqnarray}
We remark in passing that (\ref{eq_b}) means we are looking for
$\eta^H$-null geodesics (cf. section \ref{sec_MGRpostulates})
Note that this set of conditions is $O_\pc(1,3)$ invariant.
These conditions enforce that exactly one of $c$ and $d$ must be
timelike, and the other one spacelike or vanishing.
As the equations (\ref{eq_a}-\ref{eq_b}) are invariant under exchange of $c$ and $d$, we can assume without loss of generality
that $c$ be timelike. If then $d$ is vanishing, we get the solution
\begin{equation}\label{solution}
   \dot X^\mu = c^\mu, \qquad c \textrm{ unit timelike}.
\end{equation}
If $d$ is spacelike, we can (due to the $O_\pc(1,3)$-invariance of the
conditions) perform a boost such as to get
$c=\left(\gamma,0,0,0\right)$, and thus because of (\ref{eq_b}),
$d=\left(0,\delta_1,\delta_2,\delta_3\right)$, which we can rotate to
$d=\left(0,\delta,0,0\right)$ without changing $c$. Then the pseudo-rotation
\begin{equation}
  \left(\begin{array}{cccc} 1 & 0 &  0 & 0 \\
                            0 & 0 & -I & 0 \\
                            0 & I &  0 & 0 \\
                            0 & 0 &  0 & 1\end{array}\right)
\end{equation} 
will also give a solution of the form (\ref{solution}).
Using the pseudo-complex Lorentz symmetry of the equations, we find
from (\ref{solution}) all other solutions satisfying the contraints:
\begin{equation}\label{sol2}
  \dot X^\mu = c^\mu + I d^\mu.
\end{equation} 
This may look strange, but is easily understood as a consequence of the
pseudo-complex Lorentz invariance of the theory, as the transfomations
\begin{equation}
  O_\pc(1,3) \backslash O_\reals(1,3)
\end{equation}
are the analoga of symplectic transformations in classical
canonical mechanics, where \textsl{symplectic symmetry} also does not
allow for a well-defined distinction between coordinates
and momenta!\\

It is interesting to note that pseudo-complexification of the
\textsl{unconstrained} relativistic Lagrangian yields
\begin{equation}
  L_{\textrm{SR}} \equiv \dot x^\mu \dot x_\mu \longrightarrow L_{\textrm{ESR}}\equiv\dot X^\mu
  \dot X_\mu = (\dot x^\mu \dot x_\mu + \dot u^\mu \dot u_\mu) + 2 I
  \dot x^\mu \dot u_\mu,
\end{equation}
allowing us to enforce the orthogonality constraint by a
\textsl{reality condition}
\begin{equation}\label{reality}
  L_{\textrm{ESR}} \stackrel{!}{=} \bar L_{\textrm{ESR}}. 
\end{equation}

\subsection{Kaluza-Klein induced coupling to Born-Infeld theory}
The discussion of a \textsl{free} submaximally accelerated particle in
the previous section seems somewhat academic, as if there is no
external force present, the particle is \textsl{trivially}
submaximally accelerated, as it is moving along a geodesic. Still, the
exercise was worthwile as we obtained a manifestly
$O_\pc(1,3)$-invariant first order Lagrangian for a free massive
particle. Born-Infeld electrodynamics (\ref{BIaction}), having sparked
the whole investigation, provides a suitable candidate for an extended
specially relativistic external force. We now set out to construct an
\textsl{interaction term}, coupling a massive particle to Born-Infeld
electrodynamics. Conventional minimal coupling
\begin{equation}
  L_{\textrm{m.c.}} = - e \dot x^\mu A_\mu,
\end{equation}
as provisionally assumed in (\ref{prov_BI}), does not do the job, as
it is not $O_\pc(1,3)$-invariant. It is well-known that for a
specially relativistic point particle, coupling to the electromagnetic
field can be achieved using the Kaluza-Klein approach
\begin{equation}\label{KKcoupl}
  L = - \left(\dot z - A_\mu \dot x^\mu\right)^2 + g_{\mu\nu} \dot
  x^\mu \dot x^\nu,
\end{equation}
where $(x^\mu,z) \in \reals^{1,3} \times S^1$, and the conserved
momentum conjugate to the cyclic variable $z$,
\begin{equation}
  e \equiv\frac{\partial L}{\partial \dot z} = - 2 \left(\dot z - A_\mu \dot x^\mu\right)
\end{equation} 
is interpreted as the electric charge of the particle, as
(\ref{KKcoupl}) leads to the equations of motion
\begin{equation}
  \ddot x^\mu + {\Gamma^\mu}_{\alpha\beta} \dot x^\alpha \dot x^\beta
  = e {F^\mu}_\alpha \dot x^\alpha,
\end{equation}
giving the Lorentz force law.\\
In case of small electromagnetic fields, (\ref{KKcoupl}) can be
written
\begin{equation}
  L = \left(\begin{array}{c|c} g_{\mu\nu} & A_\mu \\ \hline  A_\nu &
  -1 \end{array}\right)_{mn} dx^m dx^n \equiv g_{mn} dx^m dx^n,
\end{equation}
where $x^m \equiv (x^\mu, z)$, and the dilaton was set to $-1$ (as the
extra dimension must be \textsl{spatial} for reasons of
causality). When coupling to Maxwell theory, the restriction to
\textsl{small} electromagnetic fields is quite problematic due to the
well-known field singularities. With Born-Infeld theory providing the
external force, on the other hand, we are much better off in this
respect. For flat background spacetime geometry, it is therefore
tempting to simply \textsl{pseudo-complexify} the relativistic
Lagrangian (\ref{KKcoupl}), as this was so successful in obtaining the
extended relativistic version of the free particle. But the fact that
even for flat spacetime $g=\eta$ the Kaluza-Klein manifold is curved,
is a clear enough \textsl{caveat} and we choose to be careful and use
the full machinery of the generally curved case. Remarkably, it will
turn out that for slowly varying vector potential $A$,
pseudo-complexification would have done the job equally well.\\
In order to facilitate interpretation of the following results,
we take the burden to calculate the diagonal lift of the Kaluza-Klein
metric in induced coordinates, i.e.
\begin{equation}\label{KKdiag}
   \left(g_{mn}\right)^D = \left(\begin{array}{cc} g_{mn} + g_{ts}{\Gamma^t}_m{\Gamma^s}_n
  & {\Gamma^t}_n g_{tm} \\ {\Gamma^t}_m g_{tn} & g_{mn} \end{array}\right).
\end{equation}
The $(4+1)$-dimensional Levi-Civit\`a connection evaluates for flat
spacetime $g=\eta$ to
\begin{eqnarray}
  {\Gamma^a}_{\beta\gamma} &=& \frac{1}{2} g^{a4}D_{\beta\gamma},\\
  {\Gamma^a}_{4\gamma} &=& \frac{1}{2} g^{a\mu}F_{\mu\gamma},\\
  {\Gamma^a}_{44} &=& 0,
\end{eqnarray}
where 
\begin{eqnarray}
  F_{\mu\nu} &\equiv& A_{\mu,\nu} - A_{\nu,\mu},\\
  D_{\mu\nu} &\equiv& A_{\mu,\nu} + A_{\nu,\mu}.
\end{eqnarray}
Denote points of the tangent bundle of the Kaluza-Klein manifold 
\begin{equation}
  X^M \equiv \left(x^m, u^m\right) \equiv \left((x^\mu,z),(u^\mu,w)\right).
\end{equation}
Defining
\begin{equation}
  E_{\phi r s} \equiv D_{\nu\phi} \delta_r^\nu \delta_s^4 + F_{s\phi}
  \delta_r^4, \qquad F_{4\pi} \equiv 0 \equiv F_{44}, 
\end{equation}
we obtain 
\begin{eqnarray}
  g_{st} {\Gamma^t}_\phi &=& \frac{1}{2} E_{\phi r s} u^r,\\
  g_{st} {\Gamma^t}_4    &=& \frac{1}{2} F_{s\nu} u^\nu,
\end{eqnarray}
determining the off diagonal blocks of (\ref{KKdiag}). Further, one
finds 
\begin{eqnarray}
\left[g_{st} {\Gamma^t}_i
{\Gamma^s}_j\right]_{\stackrel{i=\phi}{j=\psi}} &=& \frac{1}{4}
g^{rp} E_{\phi s r} E_{\psi t p} u^s u^t, \\
\left[g_{st} {\Gamma^t}_i
{\Gamma^s}_j\right]_{\stackrel{i=\phi}{j=4}} &=& \frac{1}{4} g^{rt}
E_{\phi s r} F_{t\nu} u^s u^\nu,\\ 
\left[g_{st} {\Gamma^t}_i
{\Gamma^s}_j\right]_{\stackrel{i=4}{j=\psi}} &=& \frac{1}{4} g^{rs} E_{\psi t
r} F_{s\nu} u^t u^\nu,\\
\left[g_{st} {\Gamma^t}_i
{\Gamma^s}_j\right]_{\stackrel{i=4}{j=4}} &=& \frac{1}{4} g^{\sigma\mu}
F_{\sigma\nu} F_{\mu\rho} u^\nu u^\rho,
\end{eqnarray}
determining the left upper block in (\ref{KKdiag}).
Observing that $E_{\phi r \psi} = - E_{\psi r \phi}$, we get 
\begin{eqnarray}
  L &=& \left(g^D\right)_{MN} dX^M dX^N\\
    &=& g_{ij} \left(\dot x^i \dot x^j + \dot u^i \dot u^j\right)\\
    & & + \frac{1}{2} \left(D_{\nu\phi} u^\nu \dot x^\phi \dot w +
        F_{\phi\nu} u^\nu \dot z \dot u^\phi\right)\\
    & & + \frac{2}{2} \left(D_{\nu\psi} u^\nu \dot u^\psi \dot z +
        F_{\psi\nu} u^\nu \dot w \dot x^\psi\right)\\
    & & + \frac{1}{4} \left(g^{rp} E_{\phi s r} E_{\psi t p} u^s u^t
        \dot x^\phi \dot x^\psi + g^{\sigma\mu} F_{\sigma\nu} F_{\mu\rho}
        u^\nu u^\rho \dot z^2\right)\\
    & & + \frac{1}{2} g^{tr} F_{t\nu} E_{\phi s r} u^\nu u^s \dot
    x^\phi \dot z.
\end{eqnarray}
For slowly varying vector potential $A$, all terms $D$,
$F$, and hence $E$ are small, and to lowest order, we get
\begin{equation}\label{coupling0}
  L_{0} = - \left(\dot z - A_\mu \dot x^\mu\right)^2 - \left(\dot w -
  A_\mu \dot u^\mu\right)^2 + \eta_{\mu\nu} \left(\dot x^\mu \dot x^\nu
  +\dot u^\mu \dot u^\nu \right),
\end{equation}
the Lagrangian for a free extended specially relativistic particle,
plus an $O_\pc(1,3)$-invariant coupling term! This result
justifies to push the analysis for slowly varying vector potentials
$A$ a bit further: We will \textsl{redo} the calculation, now using the
\textsl{pseudo-complexification prescription} to obtain the extension of
(\ref{KKcoupl}), thus being able to take care of constraints by
imposing the reality condition (\ref{reality}).\\
Direct pseudo-complexification 
\begin{eqnarray}
  x^\mu &\longrightarrow& x^\mu + I u^\mu,\\
  z &\longrightarrow& z + I w,
\end{eqnarray}
of the Lagrangian (\ref{KKcoupl}) yields
\begin{eqnarray}\label{coupling0_constr}
L_0^{\textrm{constr}} = L_{0} + I \left\{-\left(\dot z - A_\mu \dot x^\mu\right)\left(\dot
w - A_\mu \dot u^\mu\right) + g_{\mu\nu}\dot x^\mu \dot u^\nu \right\},
\end{eqnarray}    
generating an additional pseudo-imaginary part compared with
(\ref{coupling0}), which we will come back to in a moment.\\
The conserved quantities 
\begin{eqnarray}
  e_1 &\equiv& - 2 \left(\dot z - A_\mu \dot x^\mu\right),\\
  e_2 &\equiv& - 2 \left(\dot w - A_\mu \dot u^\mu\right),
\end{eqnarray}
are easily interpreted from the equations of motion for
(\ref{coupling0_constr}),
\begin{equation}\label{eqomotion}
  \ddot X^\mu = \left(e_1 + I e_2\right) F_{\mu\nu} \left(\dot x^\mu +
  I \dot u^\mu\right).
\end{equation}
Additional to the familiar coupling of the velocity to the
electromagnetic field, controlled by $e_1$, there is now also an
\textsl{a priori}
possible coupling of the acceleration, controlled by $e_2$.
However, if we now impose the reality condition on (\ref{coupling0_constr}),
\begin{equation}
  L \stackrel{!}{=} \bar L, 
\end{equation}
in order to generate the constraints, we find
\begin{equation}\label{breakortho}
  g_{\mu\nu} \dot x^\mu \dot u^\nu = \frac{e_1 e_2}{4}.
\end{equation}
In contrast, the orthogonality condition 
\begin{equation}
 g^H\left(\dot X,\dot X\right) = 0
\end{equation}
for any orbits representing submaximally accelerated particles shows
that for electrically charged particles, we must set $e_2 \equiv 0$, i.e. there is no such 'acceleration
coupling' possible in the framework of extended relativity.
So from (\ref{eqomotion}) we obtain the equation of motion for a
submaximally accelerated particle coupled to a (Born-Infeld)
electromagnetic field as
\begin{equation}
  \ddot X^\mu = e {F^\mu}_\nu \dot X^\nu,\qquad (e \textrm{ being the electric
  charge})
\end{equation}
which, of course, is roughly speaking just two 'copies' of the Lorentz
force law in the real and pseudo-imaginary part. This leads to the
remarkable conclusion that the Lorentz force law as such is also
extended specially relativistic, without any modification!

\section{Quantization}\label{sec_quant}
Classical Hamiltonian mechanics is the study of (non-relativistic)
phase space functions and their evolution in time determined by the
Hamiltonian $H$ of the system at hand. Classical phase space carries a
complex structure $\omega^{ij}$, satisfying
\begin{equation}
  {\omega^a}_j {\omega^j}_b = - \delta^a_b.
\end{equation}
It is well-known that defining the Poisson bracket
\begin{equation}
  \left[F,G\right]_{\textrm{P.B.}} \equiv \omega^{ij} \partial_i F\,
  \partial_j G,
\end{equation}
where indices run over all phase space axes, the time evolution of a
classical observable $F=F(X)$ is determined by
\begin{equation}
  \left[H,F\right] = \frac{dF}{dt}.
\end{equation}  
In particular, the classical Poisson bracket relations for the phase
space coordinates follow from Hamilton's equations as
\begin{eqnarray}
  \left[q^m, q^n \right]_{\textrm{P.B.}} &=& 0,\\
  \left[q^m, p_n \right]_{\textrm{P.B.}} &=& \delta^m_n,\\
  \left[p_m, p_n \right]_{\textrm{P.B.}} &=& 0.
\end{eqnarray}
Wigner's prescription for the transition to quantum mechanics simply
consists of a one-parameter ($\hbar$) \textsl{deformation} of the classical
Poisson bracket. Notably, it does \textsl{not} involve promotion of
classical phase space functions to operators acting on a Hilbert
space, but is nevertheless a fully equivalent description, as shown in
e.g. \cite{Zachos}.\\
Defining the \textsl{star product}
\begin{equation}\label{starp}
  * \equiv \exp\left(\frac{i \hbar}{2} \stackrel{\leftarrow}{\partial_i}
\omega^{ij} \stackrel{\rightarrow}{\partial_j}\right),
\end{equation}
the \textsl{Moyal bracket}
\begin{equation}\label{Moyal}
  \left[F,G\right]_{\textrm{M.B.}} \equiv \frac{1}{i\hbar} \left(F * G
  - G * F\right)
\end{equation}
provides the desired deformation, as can be seen from its expansion
in $\hbar$:
\begin{equation}
  \left[F,G\right]_{\textrm{M.B.}} = \left[F,G\right]_{\textrm{P.B.}}
  + \mathcal{O}(\hbar). 
\end{equation}
Note that all contributions from \textsl{even powers} of the star product (\ref{starp})
\textsl{cancel} in the definition of the Moyal bracket (\ref{Moyal}). 
In the Wigner formalism, the anti-symmetric Moyal bracket plays a
r\^ole analogous to the commutation relations in the operator
formalism. Hence the \textsl{commutation relations} in the quantum
theory stem from the \textsl{complex structure} of classical phase
space.\\

Now consider extended special relativity and the associated
phase space
\begin{equation}
  \left(\pc^4, \eta\right) \cong \left(T\reals^4, \eta^H, \eta^D\right),
\end{equation}
featuring the \textsl{pseudo-complex structure} $\eta^H$, satisfying
\begin{equation}
   {\left(\eta^H\right)^a}_m {\left(\eta^H\right)^m}_b = + \delta^a_b.
\end{equation}
In view of what we found above, this \textsl{apparently does not} give rise to commutation
relations. However, define the
\textsl{moon product}
\begin{equation}
  \odot \equiv \sinh\left(\frac{i\hbar}{2}  \stackrel{\leftarrow}{\partial_i}
\left(n^H\right)^{ij} \stackrel{\rightarrow}{\partial_j} \right),
\end{equation}
dropping all \textsl{even} powers compared to the star product
(\ref{starp}). This enables us to define an \textsl{anti-Moyal
bracket}
\begin{equation}
  \left\{F,G\right\}_{\textrm{M.B.}} \equiv \frac{1}{i\hbar} \left(F
  \odot G + G \odot F\right),
\end{equation}
whose expansion in $\hbar$ yields
\begin{equation}
  \left\{F,G\right\}_{\textrm{M.B.}} = \left(n^H\right)^{ij}
  \partial_i F \, \partial_j G + \mathcal{O}(\hbar).
\end{equation}
The lowest order term of this expansion can be interpreted as an
\textsl{anti-Poisson bracket}.
We conclude that the \textsl{pseudo-complex structure} of phase space in
extended relativity gives rise to \textsl{anti-commutation relations}
after transition to the quantized theory. Spinors being more
fundamental than tensors, one can construct commuting tensors from
anticommuting spinors, but not the other way around.\\[5pt]
In this sense, the pseudo-complex structure proves to \textsl{embrace} the structures
which are always thought of as being intimately related to a complex
structure on classical phase space.\\

\section{Conclusion}
The dynamical symmetries of Born-Infeld theory associated with the
maximal acceleration of particles coupled to it can be encoded in a
pseudo-complex geometry of the tangent bundle of
spacetime.\\
Considering the theory on this space, we classify these symmetries
then as \textsl{kinematical}, and indeed the corresponding symmetry
group preserving the geometrical structures contains transformations
to uniformly accelerated frames and a relativistic analog of the
classical symplectic transformations.\\
A particularly concise prescription for the implementation of an upper
bound on worldline accelerations is the
\textsl{pseudo-complexification} of real Minkowski vector space to a
metric module. Iteration of the pseudo-complexification process, as
mathematically developed in section \ref{sec_pciterat}, can be shown
to put upper bounds on arbitrarily many higher worldline derivatives beyond
acceleration.\\
The applicability of this prescription likewise to vector spaces, algebras and
groups acting on them on one hand, and merely Lorentz invariant
Lagrangians on the other, in order to translate them to their extended
relativistic counterparts, makes the formalism so worthwile.\\
In the generally curved case, the pseudo-complex structure surfaces
again manifestly in the adapted frames. For the purposes of this
essay, however, we made use of the numerous results from differential
geometry of tangent bundles and thus illustrated the \textsl{bimetric
real manifold} point of view.\\
The lift of the Einstein field equations, and notably the recasting of
Lagrangians with second order derivatives ('dynamically enforcing'
maximal acceleration) into first order form, were only possible
because of the identification and use of the pseudo-complex phase
spacetime structure.\\
The lifted Kaluza-Klein mechanism proved successful in generating an extended
 specially relativistic coupling of an electrically charged
particle to Born-Infeld theory, making essential use of the relevant
constructions for generally curved pseudo-complex manifolds.\\
The pseudo-complex structure leading to anti-commutation relations in
the transition to the associated quantum theory sheds a new light on
the 'origin' of anticommutation relations, which are more
fundamental than commutation relations, in the same sense as spinors
are more fundamental than tensors. These results, however, certainly deserve
further investigation.

\section*{Acknowledgments}
I would like to thank Gary Gibbons for very helpful discussions and
remarks on the material of this paper. I have also benefitted from
remarks by Paul Townsend and discussions with Sven Kerstan. 
This work is funded by EPSRC and Studienstiftung des deutschen Volkes. 

%\begin{noteinproof}
% A note added in proof, if there is one, should be the final text 
% before the references.
%\end{noteinproof}

\end{document}